\documentclass[journal]{IEEEtran}
\usepackage{geometry}
\ifCLASSINFOpdf

\else

\fi
\usepackage[cmex10]{amsmath}
\usepackage{amssymb}
\usepackage{cite}

\usepackage{graphicx}
\usepackage{array,color}
\usepackage{amsmath}
\usepackage{stfloats}
\usepackage{graphicx}
\usepackage{subfigure}
\usepackage{tabularx}
\usepackage{epsfig,epsf,color,balance,cite}
\usepackage{algorithmic}
\usepackage{algorithm}
\usepackage{bm}
\usepackage{textcomp}

\makeatletter

\renewcommand{\maketag@@@}[1]{\hbox{\m@th\normalsize\normalfont#1}}%

\makeatother
\allowdisplaybreaks[3]
\begin{document}
\title{Joint Beamforming Design for Double Active RIS-assisted Radar-Communication Coexistence Systems}
\author{ Mengyu Liu, Hong Ren, \IEEEmembership{Member, IEEE}, Cunhua Pan, \IEEEmembership{Senior Member, IEEE}, Boshi Wang, Zhiyuan Yu, Ruisong Weng, Kangda Zhi, Yongchao He
\thanks{M. Liu, H. Ren, C. Pan, B. Wang, Z. Yu, R. Weng, and Y. He  are with National Mobile	Communications Research Laboratory, Southeast University, Nanjing, China. (e-mail:{mengyuliu, hren, cpan, 220230800, zyyu, ruisong\_weng, heyongchao}@seu.edu.cn). K. Zhi is with the School of Electronic Engineering and Computer Science at Queen Mary University of London, London E1 4NS, U.K. (e-mail:k.zhi@qmul.ac.uk).
	
	Corresponding author: Hong Ren and Cunhua Pan.}
}

\maketitle
\vspace{-1.9cm}
\begin{abstract}
Integrated sensing and communication (ISAC) technology has been considered as one of the key candidate technologies in the next-generation wireless communication systems. However, when radar and communication equipment coexist in the same system, i.e. radar-communication coexistence (RCC), the interference from communication systems to radar can be large and cannot be ignored.  Recently, reconfigurable intelligent surface (RIS) has been introduced into RCC systems to reduce the interference. However, the ``multiplicative fading" effect introduced by passive RIS limits its performance. To tackle this issue, we consider a double active RIS-assisted RCC system, which focuses on the design of the radar's beamforming vector and the active RISs' reflecting coefficient matrices, to  maximize  the achievable data rate of the communication system. The considered system needs to meet the radar detection constraint and the power budgets at the radar and the RISs. Since the problem is non-convex, we propose an algorithm based on the penalty dual decomposition (PDD) framework. Specifically, we initially introduce auxiliary variables to reformulate the coupled variables into equation constraints and incorporate these constraints into the objective function through the PDD framework. Then, we decouple the equivalent problem into several subproblems by invoking the block coordinate descent (BCD) method. Furthermore, we employ the Lagrange dual method to alternately optimize these subproblems. Simulation results verify the effectiveness of the proposed algorithm. Furthermore, the results also show that under the same power budget,  deploying double active RISs in RCC systems can achieve higher data rate than those with single active RIS and double passive RISs.
\end{abstract} 
\begin{IEEEkeywords}
	Active reconfigurable intelligent surface (RIS), integrated sensing and communication (ISAC), radar-communication coexistence (RCC), penalty dual decomposition (PDD) algorithm.
\end{IEEEkeywords}

\IEEEpeerreviewmaketitle
\section{Introduction}
\IEEEPARstart{N}{ext-generation} wireless communication systems are expected to  provide $100$ times higher connection density compared with the fifth generation (5G) wireless communication system to support diverse communication scenarios \cite{you2021towards}. However, the existence of a large number of devices aggravates the demand for wireless spectrum resources. To satisfy the demand, integrated sensing and communication (ISAC) technology has been advocated to share the frequency bandwidth with radar systems \cite{liu2022integrated, liu2022survey}.
Specifically, this technology can integrate the communication system and radar system by sharing frequency band resources, transmission waveforms, and hardware platforms. Therefore, ISAC can effectively reduce the hardware cost and improve the spectral and energy efficiencies.

There are two main implementation approaches for ISAC \cite{liu2020joint}: the radar-communication coexistence (RCC) system and the dual-functional radar and communication (DFRC) system.  For the former, the hardware parts of the radar and the communication system are separate \cite{labib2017coexistence}, while for the latter the hardware parts of the radar and the communication system are integrated \cite{liu2020joint}.  Compared with DFRC systems, RCC systems typically refrain from altering the existing hardware structure and only consider the design of resource allocation strategies. With this in mind, RCC systems have attracted considerable interest\cite{zheng2019radar}. However, the interference between radar and communication equipment can seriously degrade the achievable data rate of the communication system due to the coexistence.
Recently, many researchers have made extensive contributions on the RCC systems to reduce the interference between communication systems and radar. By negotiating the spectrum use, the authors of \cite{li2016optimum} proposed the idea of cooperation between radar and communication systems to mitigate mutual interference while fulfilling power and rate constraints. In addition, the authors of \cite{liu2018mimo} proposed a optimization algorithm based on the gradient projection method to solve the power minimization problem by exploiting the constructive multiuser interference. 

 Recently, some works have deployed reconfigurable intelligent surface (RIS) in RCC systems to increase the degrees-of-freedom (DoFs) for designing the transmission strategy. Specifically, RISs are programmable meta-surfaces comprising multiple passive reflecting elements, each of which introduces a tunable phase shift to the incident signal. By controlling these reflecting elements, the signals reaching the intended receiver can be enhanced, while those reaching the unintended receiver can be suppressed \cite{liu2023integrated, pan2022overview}. Therefore, RISs are endowed with the ability to customize the wireless channel and thus enhance the signal transmission, which enables its extensive  application in wireless communications \cite{pan2020multicell,wu2019intelligent}. Inspired by this, some researchers have already introduced the RIS into RCC systems to eliminate the interference between the communication systems and the radar  \cite{wang2020ris, he2022ris}. The authors of \cite{wang2020ris} first proposed the concept of deploying single RIS in RCC systems for enhancing the detection probability of radar. Unlike the single RIS-assisted RCC system considered in \cite{wang2020ris}, the authors of \cite{he2022ris} deployed two RISs in RCC systems to eliminate the interference between radar and communication systems, which demonstrates that due to the RIS cooperation gain, deploying double RISs in RCC systems leads to better system performance than that of single RIS. 

 However, the gain of passive RIS is limited due to the fact that the signal reflected through the RIS suffers from double pathloss, resulting in severe attenuation at the receiver. This effect is called ``multiplicative fading'' \cite{wu2021intelligent, bjornson2019intelligent}. Fortunately, the concept of active RIS is proposed to overcome these limitations introduced by passive RIS \cite{zhang2022active,zhi2022active,zhou2023framework}. Different from passive RIS, active RIS is equipped with an integrated active amplifier on each reflective element to compensate for signal attenuation, so that the gain of RIS can be fully unleashed. Motivated by this, researchers have introduced active RIS into DFRC systems \cite{yu2023active,salem2022active}.  In  \cite{yu2023active}, an active RIS was deployed in DFRC systems to enhance radar sensing performance and communication quality. Simulation results of \cite{yu2023active} demonstrated the advantages of deploying active RIS in DFRC systems compared to passive RIS. Additionally, in \cite{salem2022active}, active RIS was applied to improve physical layer security in DFRC systems. These literature validated the superiority of active RIS over passive RIS in DFRC systems. However, they mainly focused on DFRC scenarios with a single active RIS. To the best of the authors' knowledge, the active RIS-aided RCC systems have not been studied, which is also an important application scenario for ISAC. Besides, considering the promising cooperative gain between double RISs, the investigation of double active RIS-aided RCC systems is of significant importance, and it is expected to achieve superior performance.
\begin{figure}[t]
	\centering
	\includegraphics[width=1\linewidth]{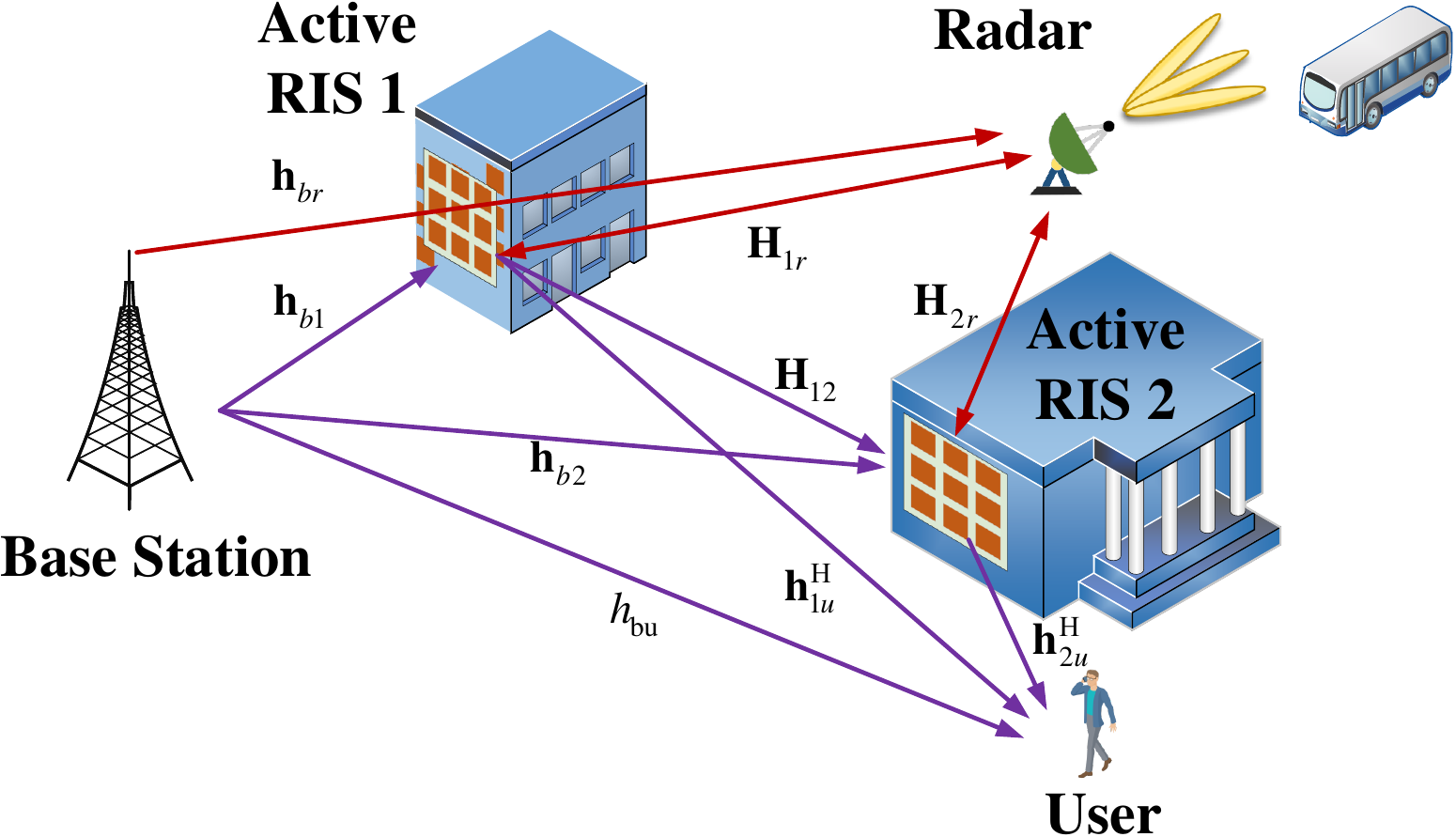}
	\caption{Double active RIS-assisted radar-communication coexistence system.}\vspace{-0.3cm}
	\label{fig1}
\end{figure}

Therefore, in this paper, we investigate a double active RIS-assisted RCC system. Due to the presence of active RIS, the amplification of thermal noise and the power constraints of active RIS need to be considered, which means that the methods proposed by \cite{wang2020ris} and \cite{he2022ris} cannot be directly applied. Meanwhile, the deployment of double RISs also introduces non-trivial challenges for the beamforming design. To tackle these challenges, we propose a low-complexity algorithm to solve the formulated optimization problem.
The main contributions of this paper are summarized as follows:
\begin{enumerate}
\item To the best of our knowledge, this is the first attempt
to investigate double active RIS-assisted RCC systems.  By optimizing the radar's beamforming vector and the active RISs' reflecting coefficient matrices, we formulate the optimization problem that maximizes the achievable data rate of the communication system subject to the radar detection constraint and the power budgets at the radar and the RISs.
\item  Then, we propose a low complexity algorithm based on the penalty dual decomposition (PDD) framework, which transforms the original problem into  a more tractable form. In the inner loop, the problem can be divided into several subproblems using the block coordinate descent (BCD) method and each subproblem can be solved by using the Lagrange dual method. In the external loop, the dual and penalty parameters are updated. By utilizing our proposed low-complexity algorithm, the computational speed for identifying the optimal solution can be significantly enhanced.
\item To draw more insights, we derive a two-loop algorithm based on the PDD framework for the double-passive-RIS-assited RCC system as a benchmark scheme. Different from active RIS, passive RIS introduces a unit-modulus constraint. To tackle this issue with low complexity, we use the  Minorize-Maximization (MM) algorithm to optimize the phase shift matrix of passive RIS. 
\item Finally, our simulation results reveal the advantages of the proposed scheme. We demonstrate that double active RISs can achieve much higher data rate than single active RIS and double passive RISs under the same power budget and the same number of reflecting elements. In addition, we verify the convergence and effectiveness of the proposed algorithm. 
\end{enumerate}

The remainder of this paper is organized as follows. In Section II, we present the system model and formulate the achievable data rate maximization problem. In Section III, a PDD-based algorithm is
 proposed to solve this problem. In Section IV, a benchmark algorithm based on passive RIS is derived. Finally, Sections V and VI provide the numerical results and conclusions, respectively.
\begin{figure*}[hb] 
	\centering 
	\hrulefill 
	\vspace*{1pt} 
	\newcounter{TempEqCnt2} 
	\setcounter{TempEqCnt2}{\value{equation}} 
	\setcounter{equation}{1} %
	\begin{align}
		\label{radar_rs}
		\nonumber		{\bf{y}}^{\rm{r}}_{k}[t] =&\underbrace {{\bf A}_{k} {\bf{x}}[t-t_0]}_{{\rm{Echo\ signal\ from\ the\ target}}} +\underbrace {({\bf{H}}_{1r}{\bm{\Theta}}_1+{{\bf{H}}_{2r}{\bm{\Theta}}_2{\bf{H}}_{12}{\bm{\Theta}}_1}) {\bf{n}}_{1}[t]+ {{\bf{H}}_{2r}{\bm{\Theta}}_2 {\bf{n}}_{2}[t]} + {\bf{n}}[t]}_{\rm{Thermal\ nosie}} \\ &+  { \underbrace {\sum\limits_{m = 1, m \ne k}^K {\bf A}_{m} {\bf{x}}[t-t_0]}_{{\rm{Interference\ from\ other\ targets}}} }+\underbrace {({\bf{h}}_{br}+{\bf{H}}_{1r}{\bm{\Theta}}_1 {\bf{h}}_{b1}+{{\bf{H}}_{2r}{\bm{\Theta}}_2 {\bf{h}}_{b2} +{\bf{H}}_{2r}{\bm{\Theta}}_2{\bf{H}}_{12}{\bm{\Theta}}_1 {\bf{h}}_{b1}})\sqrt{P^{\rm{act}}_{\rm{t}}}c[t]}_{\rm{Interference\ from\ BS\ to\ radar}},
	\end{align}
	\hrulefill 
	\vspace*{1pt} 
	\newcounter{TempEqCnt3} 
	\setcounter{TempEqCnt3}{\value{equation}} 
	\setcounter{equation}{2} %
	\begin{equation}
		\label{radar_SINR}
		{\rm{SINR}}^{\rm{r}}_{k} =  \frac{{\left| { {{\bf{d}}^{\rm{H}}_{k}}{\bf{A}}_{k}{{\bf{w}}_{k}}} \right|^2}} {\sum\limits_{m = 1, m \ne k}^K{\left| { {{\bf{d}}^{\rm{H}}_{k}}{\bf{A}}_{m}{{\bf{w}}_{k}}} \right|^2}+{P^{\rm{act}}_{\rm{t}}}{\left| { {{\bf{d}}^{\rm{H}}_{k}}{\bf q}} \right|^2}+\sigma_{1}^2{\left\|  {{\bf{d}}^{\rm{H}}_{k}}{\bf{B}} \right\|^2_2}+\sigma_{2}^2{\left\|  {{\bf{d}}^{\rm{H}}_{k}}{\bf{C}} \right\|^2_2}+\sigma^2{{\left\| { {{\bf{d}}^{\rm{H}}_{k}}} \right\|^2_2}}},
	\end{equation}
\end{figure*}

\emph{Notations}: In this paper, lowercase, boldface lowercase, and boldface uppercase letters are used to represent scalars, vectors and matrices, respectively. $\bf{I}$ and $\bf{0}$ denote an identity matrix and an all-zero vector, respectively. ${{\mathbb{ C}}^{M\times N }}$ denotes the complex matrix whose dimension is $M \times N$. ${{\mathbb{E}}}\{\cdot\}$ denotes the operation of expectation. ${\rm{diag}}(\cdot)$ denotes the operation of diagonalization. Also, ${\rm{blkdiag}}(\cdot)$ denotes a block diagonal matrix. $|a| $ and ${{\rm{Re}}(a)}$ denote the absolute value, the real value of the complex number $a$. ${\| \mathbf{a}\|}_2$ is the 2-norm of vector $\mathbf{a}$.  For matrix $\mathbf{A}$, ${\rm{Tr}}\left( {\bf{A}} \right)$ and  $\|{\bf{A}}\|_{{F}}$ denote the trace operation of ${\bf{A}}$ and Frobenius norm of $\mathbf{A}$, respectively. For two matrices $\mathbf{A}$ and $\mathbf{B}$, ${\bf{A}} \odot {\bf{B}}$ is used to denote the Hadamard product. ${\cal C}{\cal N}({\bf{0}},{\bf{I}})$ denotes a vector that follows a normal distribution with zero mean and unit covariance matrix. The conjugate, transpose and Hermitian operators  are represented by ${\left( \cdot \right)^{*}}$, ${\left( \cdot \right)^{\rm{T}}}$ and ${\left(  \cdot \right)^{\rm{H}}}$, respectively.
\section{System Model and Problem Formulation}\label{system}
As shown in Fig. \ref{fig1}, we consider a double active RIS-assisted RCC system,\footnote{We assume that the base station, radar and two RISs are all connected to the same central controller, which provides the channel state information (CSI). The required CSI can be obtained by some efficient channel estimation algorithms  \cite{swindlehurst2022channel, zheng2021efficient}.} which consists of a pair of single-antenna base station (BS) and single-antenna user (UE) and a multiple-input multiple-output (MIMO) radar with 
$M$ antennas, which can detect point-like targets in $K$ directions  within one detection epoch.  Additionally, two active RISs named as active RIS 1 and active RIS 2 are deployed in this system. Without
loss of generality, we assume that the active RIS 1 and active RIS 2 are equipped with $N_1$ and $N_2$ reflecting elements, respectively. 
\subsection{Radar Model}
 We assume that the radar mainly detects potential targets in some directions. The ${K}$ directions during one detection epoch are given by ${\{\theta_k\}}$, ${\forall}k \in \mathcal{K}\triangleq \left\{1,2,\cdots,K\right\}$. $T$ is used to represent the pulse repetition interval (PRI). In addition, we assume that the the probing signal is transmitted by radar at time $t$, and the echo signal is received from the targets at time $t_0$. Therefore, the probing signal transmitted by radar can be expressed as
\newcounter{TempEqCnt1} 
\setcounter{TempEqCnt1}{\value{equation}} 
\setcounter{equation}{0} %
\begin{equation}\label{radar_ts}
	{\bf{x}}[t]=\left\{
	\begin{aligned}
		&{\bf{w}}_k s_k, &t&=(k-1)T, \forall{k} \in \mathcal{K},\\
		&\bf{0}, &t&\neq (k-1)T, \forall{k} \in \mathcal{K}.\\
	\end{aligned}
	\right.
\end{equation}
where ${s_k}$ and ${\bf{w}}_k\in \mathbb{C}^{M \times 1}$ denote the radar sensing symbols and the transmitting beamforming vector for direction  ${\theta_k}, \forall{k} \in \mathcal{K}$, respectively.  Specifically, we assume that ${{{s}}_{k}}$ is an independent Gaussian random symbol with zero mean and unit signal power, i.e., $\mathbb{E}\left[ {{{{s}}_{i}}{{{{s}}_{j}^*} }} \right] = {0}, i \neq j $, and $\mathbb{E}\left[ {{{{s}}_{k}}{{{{s}}_{k}^*} }} \right] = {1}$.

Let ${\bf{h}}_{br}\in \mathbb{C}^{M \times 1}$, ${\bf{h}}_{b1} \in \mathbb{C}^{N_1 \times 1}$, ${\bf{H}}_{1r}\in \mathbb{C}^{M \times N_1}$, and ${\bf{H}}_{2r}\in \mathbb{C}^{M \times N_2}$ denote the wireless channel for BS $\rightarrow$ radar BS $\rightarrow$ active RIS 1, active RIS 1 $\rightarrow$ radar, and active RIS2 $\rightarrow$ radar, respectively. Specifically, $\rho_{i,n_i} e^{\varphi_{i,n_i}}$ denotes the reflecting coefficient of the $n$-th reflecting element of the $i$-th active RIS, where $\rho_{i,n_i} > 1$ and  $\varphi_{i,n_i} \in [0,2\pi]$. Then, the reflecting coefficient matrix of the $i$-th active RIS is $\boldsymbol{\Theta}_i=\rm{diag}({\boldsymbol{\phi}_i})$, where ${\boldsymbol{\phi}_i} \triangleq [\rho_{i,1_i} e^{\varphi_{i,1_i}}, \cdots, \rho_{i,N_i} e^{\varphi_{i,N_i}}]^{\rm{T}}, i \in \left\{1, 2\right\}$.

The signal received by radar from the target in direction $\theta_k$ can be expressed as (\ref{radar_rs}), as shown at the bottom of this page,
where ${\bf{A}}_{k}$ $\triangleq \alpha_{k} {\bf{a}}(\theta_k){\bf{a}}^{\rm{H}}(\theta_k) \in \mathbb{C}^{M \times M} $ is the target response matrix for the direction $\theta_k$. $\alpha_{k}$ denotes the pathloss factor of the target from direction $\theta_k$ and ${\bf{a}}(\theta_k) \triangleq {\left[ {1,{e^{j\frac{{2\pi d}}{\lambda }\sin \theta_k}}, \cdots ,{e^{j\frac{{2\pi d}}{\lambda }({M} - 1)\sin \theta_k}}} \right]^{\rm{T}}} $ is the array response vector of the radar antennas, where $d$ and $\lambda$ denotes the antenna spacing and the signal wavelength, respectively. Moreover, $c[t]$ and ${P^{{\rm{act}}}_{{\rm{t}}}}$ denote the communication signal from BS and the transmitted power at BS, ${\bf{n}}[t]$ denote the thermal noise at the radar, which follows the distributions of ${\bf{n}}[t]\sim{\cal C}{\cal N}\left( {\bf{0}},\sigma^2{{\bf{I}}_M} \right)$.  

Furthermore, the received beamforming vector ${\bf{d}}^{\rm{H}}_k \in \mathbb{C}^{1 \times M}$ is used to detect the echo signal from the direction $\theta_k$. Thus, by defining the equivalent channel matrices ${\bf{B}}\triangleq {\bf{H}}_{2r} {\bm{\Theta}_2}{\bf{H}}_{12} {\bm{\Theta}_1}+ {\bf{H}}_{1r} {\bm{\Theta}_1} \in \mathbb{C}^{M \times N_1}$, ${\bf{C}} \triangleq {\bf{H}}_{2r} {\bm{\Theta}_2} \in \mathbb{C}^{M \times N_2}$, and ${\bf{q}} \triangleq {\bf{h}}_{br}+{\bf{B}} {\bf{h}}_{b1} +{\bf{C}} {\bf{h}}_{b2} \in \mathbb{C}^{M \times 1}$, the radar signal-to-interference-plus-noise ratio (SINR) in terms of the direction $\theta_k$ can be expressed as (\ref{radar_SINR}), as shown at the bottom of the previous page.
\subsection{Communication Model}

Let $h_{bu}$, ${\bf{h}}_{b1}\in \mathbb{C}^{N_1 \times 1}$, ${\bf{h}}_{b2}\in \mathbb{C}^{N_2 \times 1}$, ${\bf{H}}_{12}\in \mathbb{C}^{N_2 \times N_1}$, ${\bf{h}}^{\rm{H}}_{2u}\in \mathbb{C}^{1 \times N_2}$, ${\bf{h}}^{\rm{H}}_{1u}\in \mathbb{C}^{1 \times N_1}$ denote the wireless channel for BS $\rightarrow$ UE, BS $\rightarrow$  active RIS 1, active RIS 1 $\rightarrow$ active RIS 2, active RIS 2 $\rightarrow$ UE, and active RIS 1 $\rightarrow$ UE, respectively. Then, the received signal at the user at the time $t$ can be expressed as (\ref{BS_rs}), as shown at the bottom of this page, 
where { ${{{n}}_{0}}[t]$ and ${{\bf{n}}_{i}}[t]$ are the thermal noise at the user and the $i$-th active RIS, respectively, following the distributions of $ {{{n}}_{0}}[t] \sim{\cal C}{\cal N}\left( {{0}},\sigma_{0}^2 \right)$, and  ${{\bf{n}}_{i}}[t] \sim{\cal C}{\cal N}\left( {\bf{0}},\sigma_{i}^2{{\bf{I}}_{Ni}} \right)$.

 Note that, as shown in Fig. \ref{fig1}, the targets of interest are not located near active RIS since active RIS is usually deployed on the facade of tall buildings. In addition, when the number of radar antennas is large, the radar beam is usually narrow, making it less likely to interfere with the communication system.  Furthermore, the interference from radar to user can also be reduced using some techniques such as \cite{zheng2017adaptive,khawar2014spectrum}. Therefore, radar interference to users is likely to be small and will not be considered in this paper.
\begin{figure*}[hb] 
	\centering 
	\hrulefill 
	\vspace*{1pt} 
	\newcounter{TempEqCnt4} 
	\setcounter{TempEqCnt4}{\value{equation}} 
	\setcounter{equation}{3} %
	\begin{align}
		\label{BS_rs}
		{{\rm{y}}}^{\rm{c}}[t] = &\underbrace {{(h_{\rm{bu}}+{\bf{h}}^{\rm{H}}_{2u} {\bm{\Theta}_2}{\bf{H}}_{12} {\bm{\Theta}_1} {{\bf{h}}_{b1}}+{\bf{h}}^{\rm{H}}_{1u} {\bm{\Theta}_1} {{\bf{h}}_{b1}}+{\bf{h}}^{\rm{H}}_{2u}{\bm{\Theta}_2} {\bf{h}}_{b2})} \sqrt{P^{\rm{act}}_{\rm{t}}}c[t]}_{{\rm{Communication\ signal\ from\ BS\ to\ user}}}  \nonumber \\ &+ \underbrace{({\bf{h}}^{\rm{H}}_{2u} {\bm{\Theta}_2}{\bf{H}}_{12}{\bm{\Theta}_1} +{\bf{h}}^{\rm{H}}_{1u} {\bm{\Theta}_1}) {\bf{n}}_{1}[t]+{\bf{h}}^{\rm{H}}_{2u} {\bm{\Theta}_2}  {\bf{n}}_{2}[t]+{{n_0}}[t]}_{\rm{Thermal\ noise}}+ \underbrace{(({\bf{h}}^{\rm{H}}_{2u} {\bm{\Theta}_2}{\bf{H}}_{12} +{\bf{h}}^{\rm{H}}_{1u}) {\bm{\Theta}_1}{{{\bf{H}}_{1r}^{\rm{H}}}}+{\bf{h}}^{\rm{H}}_{2u} {\bm{\Theta}_2}{\bf{H}}^{\rm{H}}_{2r}){\bf{x}}[t]}_{\rm{Interference\ from\ radar\ to \ user}},
	\end{align}
	\hrulefill 
	\vspace*{1pt} 
	\newcounter{TempEqCnt} 
	\setcounter{TempEqCnt}{\value{equation}} 
	\setcounter{equation}{4} %
	\begin{equation}
		\label{communication_SINR}
		{\rm{SNR}}^{\rm{c}} =  \frac{{P^{\rm{act}}_{\rm{t}}}{\left| {h_{\rm{bu}}+{\bf{h}}^{\rm{H}}_{2u} {\bm{\Theta}_2}{\bf{H}}_{12} {\bm{\Theta}_1} {{\bf{h}}_{b1}}+{\bf{h}}^{\rm{H}}_{1u} {\bm{\Theta}_1} {{\bf{h}}_{b1}}+{\bf{h}}^{\rm{H}}_{2u} {\bm{\Theta}_2} {\bf{h}}_{b2}} \right|^2}} {{\sigma_{1}^2{\left\| {\bf{h}}^{\rm{H}}_{2u} {\bm{\Theta}_2}{\bf{H}}_{12} {\bm{\Theta}_1} + {\bf{h}}^{\rm{H}}_{1u}  {\bm{\Theta}_1} \right\|^2_2}+\sigma_{2}^2{\left\| {\bf{h}}^{\rm{H}}_{2u} {\bm{\Theta}_2} \right\|^2_2}}+{\sigma_0 ^2}},
	\end{equation}
\end{figure*}
Based on the above discussions, the communication signal to noise ratio (SNR) for the user during one detection epoch can be expressed as (\ref{communication_SINR}), as shown at the bottom of this page.
Then, the achievable data rate (bit/s/Hz) is given by
\newcounter{TempEqCnt5} 
\setcounter{TempEqCnt5}{\value{equation}} 
\setcounter{equation}{5} %
\begin{equation}\label{communication_data}
	{{R}}={\rm{log_2}}(1+{\mathrm{SNR}}^{\rm{c}}).
\end{equation}

  Furthermore, since the active RIS is equipped with integrated active amplifiers, the thermal noise on the active RIS will also be amplified by the same factors as the transmitted signal. Moreover, the existence of active amplifiers leads to extra power consumption \cite{zhang2022active}. Recall that, the reflecting coefficient matrix of the $i$-th active RIS is discussed in the last section, the power constraints of active RIS 1 and active RIS 2 can be expressed as follows
 \begin{subequations}
 	\begin{align}
 		&{\left\| {\bm{\Theta}}_1 {\bf{h}}_{b1} \right\|^2_2}+ \sigma^2_{1}{\left\| {\bm{\Theta}}_1  \right\|^2_{{F}}} \le P_{1},\\
 		&{\left\| {\bm{\Theta}}_2 {\bf{H}}_{12}{\bm{\Theta}}_1 {\bf{h}}_{b1} \right\|^2_2}+{\left\| {\bm{\Theta}}_2 {\bf{h}}_{b2} \right\|^2_2}+\sigma^2_{1}{\left\| {\bm{\Theta}}_2 {\bf{H}}_{12} {\bm{\Theta}}_1  \right\|^2_F} \nonumber
 		\\&+\sigma^2_{2}{\left\| {\bm{\Theta}}_2  \right\|^2_{{F}}} \le P_{2},
 	\end{align}
 \end{subequations}
where $P_{1}$ and $P_{2}$ denote the power budgets of active RIS 1 and active RIS 2, respectively.

\subsection{Problem Formulation}
In this paper, we aim to maximize the achievable data rate of the communication system by jointly optimizing the radar transmit beamforming vector $\{{\bf{w}}_k\}$, radar receive beamforming vector $\{{\bf d}_k\}$, and the reflecting coefficient matrices ${\bf{\Theta}}_1$ and ${\bf{\Theta}}_2$ of two active RISs while guaranteeing the SINR of radar for each detecting direction $\{\theta_k\}$ and power limitation of radar and two active RISs. Hence, the optimization problem can be formulated as
\begin{subequations}\label{problem}
	\begin{align}
	\mathop {\max }\limits_{\substack{\{{\bf{w}}_k\},\{{\bf{d}}_k\},\\{\bf{\Theta}}_1, {\bf{\Theta}}_2 }} \quad
		& {{R}}	\label{7a}
		\\
 \textrm{s.t.}\qquad
		&{\rm{SINR}}^{\rm{r}}_k\geq \eta,   \forall k \in \mathcal{K},\label{8b} \\
		& \sum\limits_{k = 1}^K {\left\| {{{\bf{w}}_{k}}} \right\|^2_2} \le P_{\rm{r}},\label{8c} \\ 
		& {\left\| {\bm{\Theta}}_1 {\bf{h}}_{b1} \right\|^2_2}+ \sigma^2_{1}{\left\| {\bm{\Theta}}_1  \right\|^2_{{F}}} \le P_{1},\label{8d}\\
		&  {\left\| {\bm{\Theta}}_2 {\bf{H}}_{12}{\bm{\Theta}}_1 {\bf{h}}_{b1} \right\|^2_2}+{\left\| {\bm{\Theta}}_2 {\bf{h}}_{b2} \right\|^2_2}\label{8e}  \nonumber\\&+\sigma^2_{1}{\left\| {\bm{\Theta}}_2 {\bf{H}}_{12} {\bm{\Theta}}_1  \right\|^2_2}+\sigma^2_{2}{\left\| {\bm{\Theta}}_2  \right\|^2_{{F}}} \le P_{2},
	\end{align}
\end{subequations}
where $\eta$ and $P_{\rm{r}}$ denote the SINR requirement and the power budget for the radar, respectively. Constraint (\ref{8b}) denotes the minimum SINR requirement to ensure the detection accuracy of radar. Constraint (\ref{8c}) guarantees the power budget of radar. 

However, the strong coupling among ${\bf{w}}_k,{\bf{d}}_k,{\bf{\Theta}}_1,{\bf{\Theta}}_2$ and the nonconvex properties of the objective function and constraints make Problem (\ref{problem}) particularly challenging to solve. Therefore, we will develop a low-complexity algorithm based on PDD framework to solve this problem in the following section.

\section{Joint Beamforming Design Algorithm for Active RIS}\label{algo}
 In this section, we first reformulate the original problem into a more feasible form by using fractional programming (FP) method. Then, we apply this problem into the PDD framework. After that, the BCD algorithm is used to divide the problem into several subproblems. Finnaly, we use the Lagrange dual method to solve each subproblem.
\subsection{Reformulation of the Original Problem}
Notice that the achievable data rate increases as the communication SNR increases. Therefore, in the following sections, we can replace the objective function of the original problem, i.e., $R$  with the communication SNR, i.e., $\rm{SNR}^{\rm{c}}$. 

Furthermore, in order to efficiently solve the optimization problem with a fractional objective function,  we introduce an auxiliary $x$ and apply the quadratic transformation of the FP method \cite{shen2018fractional}. As a result, we can find a lower bound for ${\rm{SNR}}^c$ which can be expressed as
\begin{align}
		{\tilde{f}}({\bf{\Theta}}_1, {\bf{\Theta}}_2, x)
		\triangleq 
		&\left|{x}\right|^2(\sigma_{1}^2{\left\| {\bf{h}}^{\rm{H}}_{2u} {\bm{\Theta}_2}{\bf{H}}_{12} {\bm{\Theta}_1}+{\bf{h}}^{\rm{H}}_{1u}  {\bm{\Theta}_1} \right\|^2_2}\nonumber\\
		& +\sigma_{2}^2{\left\| {\bf{h}}^{\rm{H}}_{2u} {\bm{\Theta}_2} \right\|^2_2}+{\sigma_0 ^2})\nonumber\\\nonumber	&-2{\rm{Re}}\{{x}^*\sqrt{P^{\rm{act}}_{\rm{t}}} (h_{\rm{bu}}+{\bf{h}}^{\rm{H}}_{1u} {\bm{\Theta}_1}{{\bf{h}}_{b1}}\\
		&+{\bf{h}}^{\rm{H}}_{2u}{\bm{\Theta}_2}{\bf{H}}_{12} {\bm{\Theta}_1} {{\bf{h}}_{b1}}+{\bf{h}}^{\rm{H}}_{2u} {\bm{\Theta}_2} {\bf{h}}_{b2})\}.
\end{align}

Then, the relationship between ${\rm{SNR}}^{\rm{c}}$ and $	f({\bf{\Theta}}_1, {\bf{\Theta}}_2, x)$ can be written as
\begin{equation}
		{\rm{SNR}}^{\rm{c}}=\mathop {\max }\limits_{x}\quad {\tilde{f}}({\bf{\Theta}}_1, {\bf{\Theta}}_2, x), 
\end{equation}
where the optimal solution $x^{\rm{opt}}$ can be represented by (\ref{kkt22}), shown at the bottom of this page.} Thus, we rewrite Problem (\ref{problem}) as follows
\begin{figure*}[hb] 
	\centering 
	\hrulefill 
	\vspace*{1pt} 
	\begin{equation}
		\label{kkt22}
		{x^{\rm{opt}}}=\frac{\sqrt{{P^{act}_{t}}} ({h_{bu}+{\bf{h}}^{\rm{H}}_{2u} {\bm{\Theta}_2}{\bf{H}}_{12} {\bm{\Theta}_1} {{\bf{h}}_{b1}}+{\bf{h}}^{\rm{H}}_{1u} {\bm{\Theta}_1} {{\bf{h}}_{b1}}+{\bf{h}}^{\rm{H}}_{2u} {\bm{\Theta}_2} {\bf{h}}_{b2}})}{{{\sigma_{1}^2({\left\| {\bf{h}}^{\rm{H}}_{2u} {\bm{\Theta}_2}{\bf{H}}_{12} {\bm{\Theta}_1}  + {\bf{h}}^{\rm{H}}_{1u}  {\bm{\Theta}_1} \right\|^2_2})+\sigma_{2}^2{\left\| {\bf{h}}^{\rm{H}}_{2u} {\bm{\Theta}_2} \right\|^2_2}}+{\sigma_0 ^2}}}
	\end{equation}
\end{figure*}
\begin{align}\label{newproblem}
\mathcal{P}_1:\quad	\mathop  {\min }\limits_{\substack{\{{\bf{w}}_k\},\{{\bf{d}}_k\},{\bf{\Theta}}_1,\\ {\bf{\Theta}}_2, {x}}} \quad&{\tilde{f}}({\bf{\Theta}}_1, {\bf{\Theta}}_2, x), 
	\\
	\textrm{s.t.}\quad&
	\textrm{(\ref{8b}), (\ref{8c}), (\ref{8d}), (\ref{8e})} . \nonumber
\end{align}

However, there is a non-linear coupling of variables in constraint (\ref{8b}), which makes Problem (\ref{newproblem}) difficult to solve. Therefore, we adopt the PDD algorithm to tackle this issue.

\subsection{The Penalty Dual Decomposition (PDD) Algorithm for the Outer Loop}
The PDD algorithm is a two-loop iterative algorithm that can effectively solve non-convex optimization problems, especially in the case of non-linear coupling of optimization variables. The inner loop of the PDD algorithm solves the lagrangian problem, while the outer loop iterates over the penalty and dual parameters until it converges \cite{shi2020penalty}.

To tackle the coupling of variables in constraint (\ref{8b}), we introduce the auxiliary variables $\{u_{k}, v_k, y_{k}, {\bf{e}}_k, {\bf{t}}_k\}$, the equality constraints associated with five auxiliary variables are as follows
\begin{subequations}
	\begin{align}\label{uk}
		& u_{k}-{ {{\bf{d}}^{\rm{H}}_{k}}{\bf{A}}_{k}{{\bf{w}}_{k}}} =0,\forall k \in \mathcal{K},\\\label{vk1}
		&v_{k}-{ {{\bf{d}}^{\rm{H}}_{k}}{\bf q}}=0,\forall k \in \mathcal{K},\\ \label{yk}
		& y_{k}-\sum\limits_{m = 1, m \ne k}^K{ {{\bf{d}}^{\rm{H}}_{k}}{\bf{A}}_{m}{{\bf{w}}_{k}}} =0,\forall k \in \mathcal{K},\\
		& {\mathbf{e}}^{\rm{H}}_{k}-{ {{\bf{d}}^{\rm{H}}_{k}}{\bf B}} ={\bf{0}},\forall k \in \mathcal{K},\label{ek}\\ 
		&{\mathbf{t}}^{\rm{H}}_{k}-{ {{\bf{d}}^{\rm{H}}_{k}}{\bf C}} ={\bf{0}},\forall k \in \mathcal{K}.\label{tk}
	\end{align}
\end{subequations}

Next, we substitute the auxiliary variables into constraint (\ref{8b}). The new constraint (\ref{8b}) can therefore be written as
\begin{align}\label{nc}
	&\eta({\left| y_{k} \right|^2}+{P^{\rm{act}}_{\rm{t}}}{\left| {v_{k}} \right|^2}+\sigma_{1}^2{\left\| {\mathbf{e}}^{\rm{H}}_{k} \right\|^2_2}\nonumber+\sigma_{2}^2{\left\|  {\mathbf{t}}^{\rm{H}}_{k}\right\|^2_2}\\&+\sigma^2{{\left\| { {{\bf{d}}^{\rm{H}}_{k}}} \right\|^2_2}})-{\left|u_{k}\right|^2} \le 0.
\end{align}

After introducing the equality constraints, we can solve Problem (\ref{newproblem}) using the PDD framework in the following section.

Based on the PDD framework, we can construct the following new objective function
	\begin{align}\label{PDDproblem}
	\mathcal{P}_2:\mathop {\min }\limits_{{\bf{\Omega}}_1}\quad  & {\tilde{f}}({\bf{\Theta}}_1, {\bf{\Theta}}_2, x)+
	\frac{1}{2\rho}\sum\limits_{k = 1}^K\left|u_{k}-{ {{\bf{d}}^{\rm{H}}_{k}}{\bf{A}}_{k}{{\bf{w}}_{k}}}+\rho \lambda_{k,1}\right|^2 \nonumber
	\\&+\frac{1}{2\rho}\sum\limits_{k = 1}^K\left|v_{k}-{ {{\bf{d}}^{\rm{H}}_{k}}{\bf q}}+\rho \lambda_{k,2}\right|^2 \nonumber \\&+\frac{1}{2\rho}\sum\limits_{k = 1}^K\left|y_{k}-{ {{\bf{d}}^{\rm{H}}_{k}}{\bf{A}}_{m}{{\bf{w}}_{k}}}+\rho \lambda_{k,3}\right|^2\nonumber
	\\&+\frac{1}{2\rho}\sum\limits_{k = 1}^K\|{\mathbf{e}}^{\rm{H}}_{k}-{ {{\bf{d}}^{\rm{H}}_{k}}{\bf B}}+\rho {\boldsymbol{\lambda}}^{\rm{T}}_{k,{\rm{e}}}\|^2_2 \nonumber
	\\&+\frac{1}{2\rho}\sum\limits_{k = 1}^K\|{\mathbf{t}}^{\rm{H}}_{k}-{ {{\bf{d}}^{\rm{H}}_{k}}{\bf C}}+\rho {\boldsymbol{\lambda}}^{\rm{T}}_{k,{\rm{t}}}\|^2_2,
	\\ \nonumber
	\textrm{s.t.}\quad
	&\textrm{(\ref{8c}), (\ref{8d}), (\ref{8e}), (\ref{nc})},
\end{align}
 where $\rho$ is a penalty parameter, and $\{ \lambda_{k,1}, \lambda_{k,2} , \lambda_{k,3}, {\boldsymbol{\lambda}}^{\rm{T}}_{k,{\rm{e}}}  {\boldsymbol{\lambda}}^{\rm{T}}_{k,{\rm{t}}} \}$ are the dual parameters. ${\bf{\Omega}}_1  \triangleq {\left\{{\bf{w}}_k,{\bf{d}}_k,{\bf{\Theta}}_1,{\bf{\Theta}}_2, x, u_{k}, v_k,y_{k},{\bf{e}}_k, {\bf{t}}_k\right\}}$.
  Note that, when $\rho \rightarrow 0$, the solutions of Problem $\mathcal{P}_2$ are equal to those of Problem $\mathcal{P}_1$. Additionally, the convergence of the PDD algorithm was proved in \cite{shi2020penalty}. 

However, since the constraint (\ref{nc}) is not convex, we need to use the concave-convex procedure (CCP) method to decompose this constraint \cite{yuille2003concave}. To this end, the constraint  (\ref{nc}) can be rewritten as
\begin{eqnarray}\label{cons}
	f({\bf{d}}_k,v_k,y_{k},{\mathbf{e}}_{k},{\mathbf{t}}_{k})-g(u_k) \le 0,  \forall k \in \mathcal{K},
\end{eqnarray}
where 
\begin{subequations}\label{cccpdef1}
\begin{align}
		&g(u_k)={\left|u_{k}\right|^2}, \\
	&f({\bf{d}}_k,v_k,y_{k},{\mathbf{e}}_{k},{\mathbf{t}}_{k})=\eta({\left| { y_{k}} \right|^2} +{P^{\rm{act}}_{\rm{t}}}{\left| {v_{k}} \right|^2}\nonumber\\&\qquad+\sigma_{1}^2{\left\|  {\mathbf{e}}^{\rm{H}}_{k} \right\|^2_2}+\sigma_{2}^2{\left\|  {\mathbf{t}}^{\rm{H}}_{k} \right\|^2_2}+\sigma^2{{\left\| { {{\bf{d}}^{\rm{H}}_{k}}} \right\|^2_2}}).
\end{align}

\end{subequations}

Then, we can approximate the convex function $g(u_k)$ in the $i$-th iteration by its first-order Taylor expansion near the point $u^{(i)}_k$, as 
\begin{eqnarray}\label{1-order}
\hat{g}(u^{(i)}_k,u_k)=2{\rm{Re}}((u^{(i)}_k)^*u_k)-{\left|u^{(i)}_k\right|^2}.
\end{eqnarray}

Therefore, the constraint (\ref{nc}) can be approximated as
\begin{eqnarray}\label{cccp}
	f({\bf{d}}_k,v_k,y_{k},{\mathbf{e}}_{k},{\mathbf{t}}_{k})-\hat{g}(u^{(i)}_k,u_k) \le 0, \forall k \in \mathcal{K}.
\end{eqnarray}

Based on the above CCP method, we can recast Problem (\ref{problem}) as
	\begin{align}\label{finalproblem}
		\mathop {\min }\limits_{{\bf{\Omega}}_1}\quad  & {\tilde{f}}({\bf{\Theta}}_1, {\bf{\Theta}}_2, x)+
		\frac{1}{2\rho}\sum\limits_{k = 1}^K\left|u_{k}-{ {{\bf{d}}^{\rm{H}}_{k}}{\bf{A}}_{k}{{\bf{w}}_{k}}}+\rho \lambda_{k,1}\right|^2 \nonumber
		\\&+\frac{1}{2\rho}\sum\limits_{k = 1}^K\left|v_{k}-{ {{\bf{d}}^{\rm{H}}_{k}}{\bf q}}+\rho \lambda_{k,2}\right|^2 \nonumber \\&+\frac{1}{2\rho}\sum\limits_{k = 1}^K\left|y_{k}-{ {{\bf{d}}^{\rm{H}}_{k}}{\bf{A}}_{m}{{\bf{w}}_{k}}}+\rho \lambda_{k,3}\right|^2\nonumber
		\\&+\frac{1}{2\rho}\sum\limits_{k = 1}^K\|{\mathbf{e}}^{\rm{H}}_{k}-{ {{\bf{d}}^{\rm{H}}_{k}}{\bf B}}+\rho {\boldsymbol{\lambda}}^{\rm{T}}_{k,{\rm{e}}}\|^2_2 \nonumber
		\\&+\frac{1}{2\rho}\sum\limits_{k = 1}^K\|{\mathbf{t}}^{\rm{H}}_{k}-{ {{\bf{d}}^{\rm{H}}_{k}}{\bf C}}+\rho {\boldsymbol{\lambda}}^{\rm{T}}_{k,{\rm{t}}}\|^2_2,
		\\ \nonumber
 \textrm{s.t.}\quad
		&\textrm{(\ref{8c}), (\ref{8d}), (\ref{8e}), (\ref{cccp})}.
	\end{align}

Note that Problem (\ref{finalproblem}) can be readily verified as a convex problem, which can be solved by using CVX \cite{grant2014cvx}. However, the computational complexity of using the CVX to solve Problem (\ref{finalproblem}) is very high. To reduce the complexity, we next solve Problem (\ref{finalproblem}) by using the Lagrange dual method.
\subsection{The BCD Algorithm for the Inner Loop}
In this subsection, we solve the augmented Lagrange (AL) problem using the BCD method \cite{bertsekas1997nonlinear}. We divide the optimization variables into eight blocks. The details of solving these subproblems are shown as follows.
\subsubsection{Optimizing the Beamforming Vector ${\bf{w}}_k$}
In this subproblem, we fix other variables to optimize ${\bf{w}}_k$. To simplify the formulation of  Problem (\ref{finalproblem}), we first define the following parameters	
\begin{subequations}\label{uk2}
\begin{align}
&{\bf{\Gamma}}_k={\bf{A}}^{\rm{H}}_{k}{{\bf{d}}_{k}}{{\bf{d}}^{\rm{H}}_{k}}{\bf{A}}_{k}+\sum\limits_{m = 1, m \ne k}^K{\bf{A}}^{\rm{H}}_{m}{{\bf{d}}_{k}}{{\bf{d}}^{\rm{H}}_{k}}{\bf{A}}_{m},\\
&{\bf{p}}_{k}=(u_k+\rho \lambda_{k,1}){\bf{A}}^{\rm{H}}_{k}{\bf{d}}_{k}+(y_{k}+\rho \lambda_{k,3}){\bf{A}}^{\rm{H}}_{m}{{\bf{d}}_{k}},\\
&{\bf{\Gamma}}={\rm{blkdiag}}([{\bf{\Gamma}}_1,\cdots,{\bf{\Gamma}}_K]), {\bf{p}}=[{\bf{p}}^{\rm{T}}_1,\cdots,{\bf{p}}^{\rm{T}}_K]^{\rm{T}}.
\end{align}
\end{subequations}
Then, Problem (\ref{finalproblem}) can be reformulated as
\begin{subequations}\label{sub1}
\begin{align}
	\mathop {\min }\limits_{{\bf{w}}} \quad 
	&{\bf{w}}^{\rm{H}}{\bf{\Gamma}}{\bf{w}}-2{\rm{Re}}\left\{{\bf{p}}^{\rm{H}}{\bf{w}}\right\},
	\\ 
	\textrm{s.t.}\qquad
	&{\bf{w}}^{\rm{H}}{\bf{w}} \leq P_{\rm{r}}, \label{con1}
\end{align}
\end{subequations}
where  ${\bf{w}}=[{\bf{w}}^{\rm{T}}_1,\cdots,{\bf{w}}^{\rm{T}}_K]^{\rm{T}}$. 
Since this subproblem (\ref{sub1}) is a quadratic constraint quadratic programming (QCQP) problem, we can use the Lagrange
multiplier method \cite{boyd2011distributed} to solve this problem. Then, the optimal solution ${\bf{w}^{\rm{opt}}}$ can be given by
\begin{eqnarray}\label{kkt4}
 {\bf{w}^{\rm{opt}}}=({\bf{\Gamma}}+\mu{\bf I})^{-1}{\bf{p}},
\end{eqnarray}
where $\mu \ge 0$ is the Lagrange multiplier which needs to satisfy the complementary slackness conditions for constraint (\ref{con1}). Specifically, we can use the bisection search method \cite{boyd2011distributed} to obtain the optimal value of $\lambda^*_0$. The algorithm to solve Problem (\ref{sub1}) is summarized in Algorithm \ref{iteda}.
\begin{algorithm}[t]
	\caption{{Bisection Search Method to Solve Problem (\ref{sub1})}}\label{iteda}
	\begin{algorithmic}[1]
		\STATE Initialize  the accuracy $\varepsilon$, the bounds $\mu^{\rm{lb}}$ and $\mu^{\rm{ub}}$;
		\REPEAT
		\IF{$\sum\limits_{k = 1}^K {\left\| {{{\bf{w}}^{\rm{opt}}}}(0) \right\|^2} \le P_{r}$}
		\STATE Terminate;
		\ELSE
		\STATE continue;
		\ENDIF
		\STATE {Compute  $\mu = {{\left( {\mu^{\rm{lb}} + \mu^{\rm{ub}}} \right)} \mathord{\left/
					{\vphantom {{\left( {{\lambda _l} + {\lambda _u}} \right)} 2}} \right.
					\kern-\nulldelimiterspace} 2}$;}
		\IF{$\sum\limits_{k = 1}^K {\left\| {{{\bf{w}}^{\textrm{opt}}}}(\mu) \right\|^2} \le P_{\rm{r}}$}
		\STATE ${\mu^{\rm{ub}}}={\mu}$;
		\ELSE
		\STATE ${\mu^{\rm{lb}}}={\mu}$;
		\ENDIF
		\UNTIL  {$\left| {\mu^{\rm{lb}} - \mu^{\rm{ub}}} \right| \le \varepsilon $}.
	\end{algorithmic}
\end{algorithm}
\subsubsection{Optimizing the Receiving Beamforming Vector ${\bf{d}}_k$}
In this block, we optimize ${\bf{d}}_k$ by fixing other variables. To simplify the representation, we first define the following parameters
\begin{subequations}\label{dk}
	\begin{align}
		\centering
		&{\bf{\beth}}_k={\bf{A}}_k {\bf{w}}_k { {\bf{w}}_k}^{\rm{H}}{\bf{A}}^{\rm{H}}_k+\sum\limits_{m = 1, m \ne k}^K{\bf{A}}_m {\bf{w}}_k{\bf{w}}^{\rm{H}}_k{\bf{A}}^{\rm{H}}_m\nonumber\\
		&\qquad+{\bf{q}}{\bf{q}}^{\rm{H}}+{\bf{B}}{\bf{B}}^{\rm{H}}++{\bf{C}}{\bf{C}}^{\rm{H}},\\
		&m=\frac{1}{\sigma^2}(\eta({\left| { y_{k}} \right|^2}+{P^{\rm{act}}_{\rm{t}}}{\left| {v_{k}} \right|^2}+\sigma_{1}^2{\left\|  {\mathbf{e}}^{\rm{H}}_{k} \right\|^2_2}\nonumber\\&\qquad+\sigma_{2}^2{\left\|  {\mathbf{t}}^{\rm{H}}_{k} \right\|^2_2})-2\text{Re}((u^{(i)}_k)^*u_k)+{\left|u^{(i)}_k\right|^2}),\\
		&{\bf{z}}_{k}=(u_k+\rho\lambda_{k,1}){\bf{A}}_k {\bf{w}}_k+(v_k+\rho\lambda_{k,2}){\bf{q}}\nonumber\\
		&\qquad+(y_{k}+\rho\lambda_{k,3}){{\bf{A}}_m} {\bf{w}}_k +{\bf{B}}({\bf{e}}^{\rm{H}}_k+\rho{\boldsymbol{\lambda}}^{\rm{T}}_{k,{\rm{e}}}) \nonumber\\
		&\qquad+{\bf{C}}({\bf{t}}^{\rm{H}}_k+\rho{\boldsymbol{\lambda}}^{\rm{T}}_{k,{\rm{t}}}).
	\end{align}
\end{subequations}

Then, Problem (\ref{finalproblem}) can be reformulated as
\begin{subequations}\label{sub22}
	\begin{align}
		\mathop {\min }\limits_{{\bf{d}}_k} \quad  
		&{\bf{d}}_k^{\rm{H}}{\bf{\beth}}_k{\bf{d}}_k-2{\rm{Re}}\left\{{\bf{z}}_k^{\rm{H}}{\bf{d}}_k\right\},
		\\ 
		\textrm{s.t.}\qquad
		&{\bf{d}}_k^{\rm{H}}{\bf{d}}_k \leq m, \forall k \in \mathcal{K}.\label{constraint24}
	\end{align}
\end{subequations}

Next, we solve this subproblem (\ref{sub22}) using the Lagrange dual method. The optimal solution ${{\bf{d}}^{\rm{opt}}_k}$ can be given by
\begin{eqnarray}\label{opt33}
	{{\bf{d}}^{\rm{opt}}_k}=({\bf{\beth}}_k+\delta_k{\bf{I}}_M)^{-1}{\bf{z}}_k,\forall k \in \mathcal{K},
\end{eqnarray}
where $\delta_k \ge 0$ is the Lagrange multiplier associated with constraint (\ref{constraint24}) which can be found by the bisection search method.

\subsubsection{Optimizing the Auxiliary Parameters}In this section, we alternately optimize the auxiliary parameters in $\left\{ u_k, v_k, y_{k}, {\bf{e}}_k, {\bf{t}}_k\right\}$.

Firstly, we optimize $\{u_k\}$ while fixing other variables. Thus, original Problem (\ref{finalproblem}) can be simplified as
	\begin{align} \label{subuk}
		\mathop {\min }\limits_{u_k} \quad  
		&\left|u_{k}-{ {{\bf{d}}^{\rm{H}}_{k}}{\bf{A}}_{k}{{\bf{w}}_{k}}}+\rho \lambda_{k,1}\right|^2,
		\\ \nonumber
		\qquad\ \textrm{s.t.}\qquad
		&\textrm{(\ref{cccp})}.
	\end{align}

For this subproblem, we can use the Lagrangian dual method. Therefore, the optimal solution ${u^{\rm{opt}}_k}$ can be given by
\begin{eqnarray}\label{opt41}
	{u^{\rm{opt}}_k}={\bf{d}}_k^{\rm{H}}{\bf{A}}_k{\bf{w}}_k-\rho\lambda_{k,1}+\mu_k u^{(i)}_k, \forall k \in \mathcal{K},
\end{eqnarray}
where $\mu_k \ge 0$ is the Lagrange multiplier associated with constraint (\ref{cccp}) which can be found by the bisection search method.
It can be observed that the methods used in optimizing the variables $\{v_k, y_{k}, {\bf{e}}_k, {\bf{t}}_k\}$ are similar to optimizing $u_k$. Therefore, we can derive the optimal solutions for the remaining variables $v_k, y_{k,m}, {\bf{e}}^{\rm{H}}_k, {\bf{t}}^{\rm{H}}_k$ as
\begin{subequations}\label{opt32}
	\begin{align}
		{v^{\rm{opt}}_k}&=\frac{{\bf{d}}_k^{\rm{H}} {\bf{q}}-\rho\lambda_{k,2}}{1+\varrho_k\eta{P^{\rm{act}}_{\rm{t}}}}, \forall k \in \mathcal{K},\\
		{y^{\rm{opt}}_{k}}&=\frac{\sum\limits_{m = 1, m \ne k}^K{\bf{d}}_k^{\rm{H}} {\bf{A}}_{m}{\bf{w}}_{k}-\rho\lambda_{k,3}}{1+\zeta_{k}\eta},\forall k \in \mathcal{K}, \\
		{{\bf{e}}^{\rm{H}}_k}&=\frac{{\bf{d}}_k^{\rm{H}} {\bf{B}}-\rho{\boldsymbol{\lambda}}^{\rm{T}}_{k,{\rm{e}}}}{1+\chi_k\eta\sigma^2_{1}}, \forall k \in \mathcal{K},\\
		{{\bf{t}}^{\rm{H}}_k}&=\frac{{\bf{d}}_k^{\rm{H}} {\bf{C}}-\rho{\boldsymbol{\lambda}}^{\rm{T}}_{k,{\rm{t}}}}{1+\varsigma_k\eta\sigma^2_{2}}, \forall k \in \mathcal{K},
	\end{align}
\end{subequations}
where $\varrho_k, \zeta_{k,m}, \chi_k, \varsigma_k   \ge 0$ are the Lagrange multipliers associated with their corresponding constraints, and the values of these multipliers can be found by the bisection search method.

\subsubsection{Optimizing the Phase Shift ${\bf{\Theta}}_1$}
We optimize ${\bf{\Theta}}_1$ while fixing the other variables. 
To simplify the formulation of  Problem (\ref{finalproblem}), we first define ${\bf{T}}_1= {x}^*\sqrt{{P^{\rm{act}}_{\rm{t}}}} {\bf{H}}^{\rm{H}}_{12}{\bm{\Theta}}^{\rm{H}}_2{\bf{h}}_{2u}{{\bf{h}}^{\rm{H}}_{b1}}$, ${\bf{F}}_1 \triangleq {x}^*\sqrt{{P^{\rm{act}}_{\rm{t}}}}{{\bf{h}}_{1u}}{{\bf{h}}^{\rm{H}}_{b1}}$, ${\bf{S}}_k=({\bf{H}}_{2r} {\bm{\Theta}_2}{\bf{H}}_{12}  +{\bf{H}}_{1r})^{\rm{H}}{\bf{d}}_k({\bf{\bf{e}}}_k+\rho{\boldsymbol{\lambda}}_{k,e})$. ${\bf{t}}_1$, ${\bf{f}}_1$, and ${\bf{s}}_k$ denote the collection of diagonal elements of matrix ${\bf{T}}_1$, ${\bf{F}}_1$ and ${\bf{S}}_k$, respectively.

By using the property ${{\rm{Tr}}}({\bm{\Theta}^{\rm{H}}_1}{\bf{D}}_1{\bm{\Theta}_1})={\bm{\phi}^{\rm{H}}_1}({\bf{D}}_1\odot{\bf{I}}_{N_1}){\bm{\phi}_1}$ and ${{\rm{Tr}}}({\bm{\Theta}_1}{\bf{T}}^{\rm{H}}_1)={\bf{t}}_1^{\rm{H}}{\bm{\phi}_1}$ \cite{zhang2017matrix}, the subproblem can be rewritten as
\begin{subequations}\label{sub6n}
	\begin{align}
		\mathop {\min }\limits_{{\bm{\phi}}_1} \quad  \label{sub6n_1}
		&{\bm{\phi}^{\rm{H}}_1}{\bm{\Xi}}_{1}{\bm{\phi}_1}-2\textrm{Re}\{{\bf{u}}^{\rm{H}}_{1}{\bm{\phi}_1}\},
		\\ 
		\qquad\ \textrm{s.t.}\qquad \label{sub6n_2}
		&f_1({\bm{\phi}}_1) \triangleq {\bm{\phi}^{\rm{H}}_1}{\bf{P}}{\bm{\phi}_1}-P_{1} \le 0 ,\\ \label{sub6n_3}
		&f_2({\bm{\phi}}_1) \triangleq {\bm{\phi}^{\rm{H}}_1}{\bf{V}}{\bm{\phi}_1}-P_{{\bm{\phi}}_1}  \le 0 ,
	\end{align}
\end{subequations}
where the parameters are defined as follows
\begin{subequations}\label{phi12}
	\begin{align}
	\centering
	{\boldsymbol{\Xi}}_{1}&=\left|{x}\right|^2\sigma_{1}^2({\bf{h}}^{\rm{H}}_{2u} {\bm{\Theta}_2}{\bf{H}}_{12} + {\bf{h}}^{\rm{H}}_{1u} )^{\rm{H}}({\bf{h}}^{\rm{H}}_{2u} {\bm{\Theta}_2}{\bf{H}}_{12} + {\bf{h}}^{\rm{H}}_{1u} )\nonumber
	\\&\quad\odot{\bf{I}}_{N_1}+\frac{1}{2\rho}\sum\limits_{k = 1}^K({\bf{b}}_k{\bf{b}}^{\rm{H}}_k+{\bf{L}}_{1k}),\\
	{\bf{P}}&={\bf{I}}_{N_1}\odot ({\bf{h}}_{b1}{\bf{h}}^{\rm{H}}_{b1}+\sigma^2_{1}{\bf{I}}_{N_1})^{\rm{T}}, \\
	{\bf{u}}_{1}&={\bf{t}}_1+{\bf{f}}_1-\frac{1}{2\rho}\sum\limits_{k = 1}^K({\rm{m}}_k{\bf{b}}_k+{\bf{s}}_k),\\
	{\bf{V}}&={\bf{H}}^{\rm{H}}_{12}{\bm{\Theta}}^{\rm{H}}_2{\bm{\Theta}}_2 {\bf{H}}_{12}\odot(({\bf{h}}_{b1}{\bf{h}}^{\rm{H}}_{b1})^{\rm{T}}+\sigma^2_{1}{\bf{I}}_{N_1}), \\ {\bf{L}}_{1k}&={\sigma^2_{1}}({{\bf{d}}^{\rm{H}}_{k}}{\bf{H}}_{2r} {\bm{\Theta}_2}{\bf{H}}_{12}+{{\bf{d}}^{\rm{H}}_{k}} {\bf{H}}_{1r})^{\rm{H}}\nonumber \\
	&\quad\times({{\bf{d}}^{\rm{H}}_{k}}{\bf{H}}_{2r} {\bm{\Theta}_2}{\bf{H}}_{12} +{{\bf{d}}^{\rm{H}}_{k}} {\bf{H}}_{1r})\odot{\bf{I}}_{N_1}, \\ P_{{\boldsymbol{\phi}}_1}&=P_{2}-{\left\| {\boldsymbol{\Theta}}_2 {\bf{h}}_{b2} \right\|^2_2}-\sigma^2_{2}{\left\| {\boldsymbol{\Theta}}_2 \right\|^2_2},  \\
	{\bf{b}}^{\rm{H}}_k&={\bf{d}}^{\rm{H}}_k({\bf{H}}_{2r} {\bm{\Theta}_2}{\bf{H}}_{12}  +{\bf{H}}_{1r}){\rm{diag}}({\bf h}_{b1}),\\
	{\rm{m}}_k&={\bf{d}}^{\rm{H}}_k({\bf h}_{br}+{\bf H}_{2r}{\bm \Theta}_2{\bf h}_{b2})-v_k-\rho\lambda_{k,2}.\
\end{align}
\end{subequations}

Because ${\boldsymbol{\Xi}}_{1}$, ${\bf{P}}$ and ${{\bf{V}}}$  are semidefinite matrices, subproblem (\ref{sub6n}) is convex. Thus, we can still use the Lagrange dual method to obtain the optimal solution of ${\bm{\phi}}^{\rm{opt}}_1$, given by
\begin{eqnarray}\label{kkt62}
	{\bm{\phi}}^{\rm{opt}}_1=({\bm{\Xi}}_1+\kappa_1{\bf{P}}+\kappa_2{\bf{V}})^{-1}{\bf{u}}_1,
\end{eqnarray}
where $\kappa_1, \kappa_2 \ge 0$ are the Lagrange  multipliers associated with constraints,
\begin{algorithm}[t]
	\caption{{Ellipsoid Method to Solve Problem (\ref{sub6n}) }}\label{iteda2}
	\begin{algorithmic}[1]
		\STATE Initial the iteration number $t=0$, the ellipsoid $({\bm{\Upsilon}}^{(0)},\kappa^{(0)}_k)$, the accuracy $\varepsilon$. $f_{{\rm{obj}}}(\kappa^{(0)}_k)$ and $f_i({\bm{\phi}}^{(t)}_1)$ denote the value of objective function (\ref{sub6n_1}) and the value of constraints (\ref{sub6n}), $ i\in \{1, 2\},  k\in \{ 1, 2\}$.
		\REPEAT 
		\IF{$\kappa^{(k)}_k < 0 $}
		\STATE Compute the subgradient $g^{(t)} = \partial f_{{\rm{obj}}}(\kappa^{(t)}_k)$.
		\STATE Normalized the subgradient $\tilde{g}^{(t)}$.
		\STATE ${{\alpha}} = {\frac{f_i({\bm{\phi}}^{(t)}_1)}{{\sqrt{{\bf{g}}^{\rm{T}}{\bm{\Upsilon}}{\bf{g}}}}}}$;
		\ELSE 
		\STATE Update the upper bound $f_{\rm{best}} (\kappa^{(t)}_k)$
		\STATE Compute the subgradient $g^{(t)} = \partial f_{{\rm{obj}}}(\kappa^{(t)}_k)$;
		\STATE Normalize the subgradient $\tilde{g}^{(t)}$;
		\STATE  ${{\alpha}} = {\frac{f_{{\rm{obj}}}(\kappa^{(t)}_k)-f_{\rm{best}} (\kappa^{(t)}_k)}{{\sqrt{{\bf{g}}^{\rm{T}}{\bm{\Upsilon}}{\bf{g}}}}}}$;
		\ENDIF
		\STATE   $\kappa^{(t+1)}_k=\kappa^{(t)}_k-\frac{1+2\alpha}{3}{\bm{\Upsilon}}^{(t)}\tilde{g}^{(t)}$;
		\STATE ${\bm{\Upsilon}}^{(t+1)}= \frac{4}{3}(1-\alpha^2)({\bm{\Upsilon}}^{(t)}-\frac{2(1+2\alpha)}{3(1+\alpha)}{\bm{\Upsilon}}^{(t)}\tilde{g}^{(t)}$ $(\tilde{g}^{(t)})^{\rm{T}}{\bm{\Upsilon}}^{(t)}),$
		\STATE $t=t+1$ ;
		\STATE Calculate the value of $f_{{\rm{obj}}}$;
		\UNTIL  $\frac{\left|f_{{\rm{obj}}}(\kappa^{(t+1)}_k)-f_{{\rm{obj}}}(\kappa^{(t)}_k)\right|}{f_{{\rm{obj}}}(\kappa^{(t)}_k)} \le  \varepsilon $.
	\end{algorithmic}
\end{algorithm}
which can be determined by the ellipsoid method \cite{boydEE392}. The overall algorithm is summarized in Algorithm \ref{iteda2}.

\subsubsection{Optimizing the Phase Shift ${\bf{\Theta}}_2$}
We optimize ${\bf{\Theta}}_2$ while fixing the other variables. 
Firstly, we define ${\bf{T}}_2 \triangleq {x}^*\sqrt{{P^{\rm{act}}_{\rm{t}}}} {\bf{h}}_{2u}{{\bf{h}}^{\rm{H}}_{b1}}{\bm{\Theta}}^{\rm{H}}_1 {\bf{H}}^{\rm{H}}_{12}$, ${\bf{F}}_2 \triangleq {x}^*\sqrt{{P^{\rm{act}}_{\rm{t}}}}{{\bf{h}}_{2u}}{{\bf{h}}^{\rm{H}}_{b2}}$, ${\bf{M}}_1 \triangleq \sigma_{1}^2\left|{x}\right|^2{\bf{h}}_{2u}{\bf{h}}^{\rm{H}}_{1u}{\bm{\Theta}_1}{\bm{\Theta}^{\rm{H}}_1}{\bf{H}}^{\rm{H}}_{12}$, ${\bf{M}}_{2k} \triangleq {\bf{H}}^{\rm{H}}_{2r}{\bf{d}}_{k}{\bf{d}}^{\rm{H}}_{k}{\bf{H}}_{1r}
{\bm{\Theta}_1}{\bm{\Theta}^{\rm{H}}_1}{\bf{H}}^{\rm{H}}_{12}$, ${\bf{G}}_{k} \triangleq {\bf{H}}^{\rm{H}}_{2r}{\bf{d}}_{k}({\bf{\bf{t}}}_k+\rho{\boldsymbol{\lambda}}_{k,{\rm{t}}})^{\rm{H}}$, and ${\bf{Q}}_{k} \triangleq {\bf{H}}^{\rm{H}}_{2r}{\bf{d}}_{k}({\bf{\bf{e}}}_k+\rho{\boldsymbol{\lambda}}_{k,{\rm{e}}})^{\rm{H}}{\bf{H}}^{\rm{H}}_{12}{\bm{\Theta}^{\rm{H}}_1}$. 
${\bf{t}}_2$, $ {\bf{f}}_2$, ${\bf{m}}_1$, ${\bf{m}}_{2k}$, ${\bf{g}}_{k}$, and ${\bf{q}}_{k}$ are  the collection of diagonal elements of matrix ${\bf{T}}_2$, ${\bf{F}}_2$, ${\bf{M}}_1$, ${\bf{M}}_{2k}$, ${\bf{G}}_{k} $, and ${\bf{Q}}_{k} $, respectively. 

Then, by using the similar method used to optimize ${\bm{\Theta}}_1$, this subproblem can be formulated as
\begin{subequations}\label{sub7n}
	\begin{align}
		\mathop {\min }\limits_{{\bm{\phi}}_2} \quad  \label{sub7n_1}
		&{\bm{\phi}^{\rm{H}}_2}{\bm{\Xi}}_{2}{\bm{\phi}_2}-2\textrm{Re}\{{\bf{u}}^{\rm{H}}_{2}{\bm{\phi}_2}\}
		\\ 
 \textrm{s.t.}\quad 
		&{\bm{\phi}^{\rm{H}}_2}{\bf{Z}}{\bm{\phi}_2}  \le P_{2},
	\end{align}
\end{subequations}
where the parameter are defined as follows
\begin{subequations}\label{phi22}
\begin{align}
		\centering
	&{\boldsymbol{\Xi}}_{2}=\left|{x}\right|^2{\bf{h}}_{2u}{\bf{h}}^{\rm{H}}_{2u}\odot(\sigma^2_{1}({{\bf{H}}_{12}{\bm{\Theta}_1}{\bm{\Theta}^{\rm{H}}_1}{\bf{H}}^{\rm{H}}_{12}})^{\rm{T}}+\sigma^2_{2}{\bf{I}}_{N_2})\nonumber\\&\qquad+\frac{1}{2\rho}(\sum\limits_{k = 1}^K{\bf{r}}_k{\bf{r}}^{\rm{H}}_k+{\bf{L}}_{2k}),\\
	&{\bf{Z}}={\bf{I}}_{N_2}\odot (({\bf{H}}_{12}{\bm{\Theta}}_1 {\bf{h}}_{b1}{\bf{h}}^{\rm{H}}_{b1}{\bm{\Theta}}^{\rm{H}}_1 {\bf{H}}^{\rm{H}}_{12})^{\rm{T}}\nonumber\\&\qquad+({\bf{h}}_{b2}{\bf{h}}^{\rm{H}}_{b2}+\sigma^2_{2}{\bf{I}}_{N_2})^{\rm{T}} +\sigma^2_{1}({{\bf{H}}_{12}{\bm{\Theta}_1}{\bm{\Theta}^{\rm{H}}_1}{\bf{H}}^{\rm{H}}_{12}})^{\rm{T}}),\\
	&{\bf{L}}_{2k}=({\bf{H}}^{\rm{H}}_{2r}{\bf{d}}_k{\bf{d}}^{\rm{H}}_k{\bf{H}}_{2r})\odot(({{\bf{H}}_{12}{\bm{\Theta}_1}{\bm{\Theta}^{\rm{H}}_1}{\bf{H}}^{\rm{H}}_{12}})^{\rm{T}}+{\bf{I}}_{N_2}),\\ 
	&{\bf{u}}_{2}={\bf{t}}_2+{\bf{f}}_2+{\bf{m}}_1-\frac{1}{2\rho}\sum\limits_{k = 1}^K({\rm{n}}_k{\bf{r}}_k+{\bf{m}}_{2k}+{\bf{g}}_{k}+{\bf{q}}_{k}),\\
	&{\bf{r}}^{\rm{H}}_k={\bf{d}}^{\rm{H}}_k{\bf{H}}_{2r}( {\rm{diag}}({\bf{H}}_{12}{\bm{\Theta}_1}{\bf{h}}_{b1})  +{\rm{diag}}({\bf{h}}_{b2})),\\
	&{\rm{n}}_k={\bf{d}}^{\rm{H}}_k({\bf{h}}_{br}+{\bf{H}}_{1r}{\bm{\Theta}}_{1}{\bf{h}}_{b1})-v_k-\rho\lambda_{k,2}.
\end{align}
\end{subequations}

Then, we can get the optimal solution of ${\bm{\phi}}^{\rm{opt}}_2$, given by
\begin{equation}\label{kkt162}
	{\bm{\phi}}^{\rm{opt}}_2=({\bm{\Xi}}_{2}+\xi{\bf{Z}})^{-1} {\bf{u}}_2,
\end{equation}
where $\xi \ge 0$ is the Lagrange multiplier associated with constraints which can be determined by the bisection search method.
\subsection{The Overall Algorithm for Solving the Optimization Problem}
\begin{algorithm}[t]
	\caption{Low-complexity algorithm based on PDD framework }\label{algoall}
	\begin{algorithmic}[1]
		\STATE Initial the iteration number $t=0$, the
		tolerance of accuracy $\varepsilon$, $\rho^{(0)} > 0$, $\sigma >0$, $0<c<1$, feasible ${\bf{\Omega}}_1^{(0)}$. Calculate the objective function value of Problem (\ref{finalproblem}), denoted as ${\mathcal{P}}({\bf{\Omega}}_1^{(0)},\rho^{(0)},{\bm{\lambda}}^{(0)})$, and the achievable data rate in (\ref{communication_data}), denoted as $R^{(0)}$. ${\bm{\lambda}}^{(0)}$ is the collection of dual parameters $\{\lambda^{(0)}_{k,2}, \lambda^{(0)}_{k,m,3}, {\bm{\lambda}}^{\rm{T}(0)}_{k,{\rm{e}}}, {\bm{\lambda}}^{\rm{T}(0)}_{k,{\rm{t}}}
		\}$.  ${{h}}^{(0)}={\rm{max}}\{ |u_{k}-{ {{\bf{d}}^{\rm{H}}_{k}}{\bf{A}}_{k}{{\bf{w}}_{k}}}|, 
			|v_{k}-{ {{\bf{d}}^{\rm{H}}_{k}}{\bf q}}|,
			|y_{k}-\sum\limits_{m = 1, m \ne k}^K{ {{\bf{d}}^{\rm{H}}_{k}}{\bf{A}}_{m}{{\bf{w}}_{k}}}|,
			| {\mathbf{e}}^{\rm{H}}_{k}-{ {{\bf{d}}^{\rm{H}}_{k}}{\bf B}} |,
			|{\mathbf{t}}^{\rm{H}}_{k}-{ {{\bf{d}}^{\rm{H}}_{k}}{\bf C}} |\}$ denote the maximum value of constraint violation between (\ref{uk}), (\ref{vk1}), (\ref{yk}), (\ref{ek}) and (\ref{tk}).
		\REPEAT 
		\REPEAT
		\STATE By fixing the other variables, calculate the optimal value of ${\bf{w}}^{(t+1)}_k$ in (\ref{kkt4});
		\STATE By fixing the other variables, calculate the optimal value of ${\bf{d}}_k^{(t+1)}$ in (\ref{opt33});
		\STATE By fixing the other variable, alternately calculate the optimal value of $\{x^{(t+1)}, u_{k}^{(t+1)}, v_k^{(t+1)},y_{k}^{(t+1)},{\bf{e}}_k^{(t+1)}, {\bf{t}}_k^{(t+1)}\} $ in (\ref{opt41}) and (\ref{opt32}), respectively;
		\STATE By fixing the other variable, alternately calculate the optimal value of ${\bf{\Theta}}^{(t+1)}_1$, ${\bf{\Theta}}^{(t+1)}_2$ in (\ref{kkt62}) and (\ref{kkt162}), respectively;
		\STATE Calculate the objective function value: ${\mathcal{P}}({\bf{\Omega}}_1^{(t+1)},\rho^{(t+1)},{\bm{\lambda}}^{t+1})$;
		\UNTIL${\frac{\left|{\mathcal{P}}({\bf{\Omega}}_1^{(t+1)},\rho^{(t+1)},{\bm{\lambda}}^{(t+1)})-{\mathcal{P}}({\bf{\Omega}}_1^{(t)},\rho^{(t)},{\bm{\lambda}}^{(t)})\right|}{{\mathcal{P}}({\bf{\Omega}}_1^{(t)},\rho^{(t)},{\bm{\lambda}}^{(t)})} \le  \varepsilon}$
		\STATE Compute the maximum value of constraint violation ${{h}}^{(t+1)}$ ;
		\STATE Compute the achievable data rate ${{R}}^{(t+1)}$ ;
		\IF{${{h}}^{(t+1)} \le \sigma $}
		\STATE ${\bm{\lambda}}^{(t+1)}={\bm{\lambda}}^{(t)}+\frac{1}{\rho^{(t)}}{\bf{h}}^{(t+1)}$,
		$ {\rho^{(t+1)}} ={\rho^{(t)}} $;
		\ELSE 
		\STATE   $ {\bm{\lambda}}^{(t+1)} ={\bm{\lambda}}^{(t)} $, $ {\rho^{(t+1)}} =c{\rho^{(t)}} $;
		\ENDIF
		\STATE $t=t+1$ ;
	
		\UNTIL $\frac{\left|{{R}}^{(t+1)}-{{R}}^{(t)}\right|}{{{R}}^{(t)}} \le  \varepsilon $
	\end{algorithmic}
\end{algorithm}
As shown in Algorithm \ref{algoall}, there are two loops in the PDD framework. For the inner loop of the PDD framework, we use the BCD method to divide the original problem into several subproblems. Then, we solve the subproblems using the Lagrange dual method with the bisection search method and the ellipsoid method. For the outer loop, we update the dual variables and penalty factors according to the value of constraint violation.

The computational complexity of the proposed PDD-based algorithm can be represented by
 \begin{equation}
 	\mathcal{O}({{I}_{\rm{outer}}}{{I}_{\rm{inner}}}(o_1 +o_2 +o_3+o_4+o_5)),
 \end{equation}
where $o_1=({{M}^{3}}+{{M}^{2}}+(K^3{{M}^{3}}+K^2{{M}^{2}})log\frac{{{I}_{0}}}{\varepsilon })$, $o_2=K({{M}^{2}}({{N}_{1}}+{{N}_{2}})+log\frac{{{I}_{0}}}{\varepsilon }({M^3+M^2}))$ , $o_3=Klog\frac{{{I}_{0}}}{\varepsilon }({{M}^{2}}+MN_1+MN_2)$,  $o_4=(N_1N_2+MN_2+MN_1+N^2_1+N^2_2+({{N}^3_{1}}+N^2_1)log\frac{RG}{\varepsilon })$, and $o_5=(M^2N_2+N_1^2N_2+N_1N_2^2+({{N}^3_{2}}+N^2_2)log\frac{RG}{\varepsilon })$ denote the computational complexity of optimizing $\{{\bf{w}}_k\}$, $\{{\bf{d}}_k\}$, $\{{u}_k,{v}_k, {y}_{k}, {\bf{e}}_k, {\bf{t}}_k\}$, ${\bf{\Theta}}_1$, and ${\bf{\Theta}}_2$, respectively. ${{I}_{\rm{outer}}}$ and ${{I}_{\rm{inner}}}$ denote  the number of outer loop iterations and  inner loop iterations, respectively. ${{I}_{0}}$ is the initial interval of the bisection-search method. $G$ and $R$ denote maximum length of the sub-gradients and the length of the semi-axes on the initial ellipsoid.

\section{Joint Beamforming Design Algorithm for Passive RIS}\label{algopas}
In this section, we consider the double passive RIS-assisted RCC system to provide a comparison benchmark scheme for double active RIS-assisted RCC systems. Unlike active RIS, passive RIS does not amplify the thermal noise, at the cost of severe signal attenuation. In addition, for each element in the phase shift matrix of the passive RIS, we have unit-modulus constraints, i.e., $\left|{\bm{\phi}_1}(n)\right|=1, n = 1, \cdots, N_1, \left|{\bm{\phi}_2}(n)\right|=1, n = 1, \cdots, N_2$.
Therefore, the original Problem (\ref{finalproblem}) for double passive RISs can be formulated as
\begin{subequations}\label{passive_problem}
	\begin{align}
		\mathop {\min }\limits_{{\bf{\Omega}}_2} \quad &-\frac{P^{\rm{pas}}_{\rm{t}}}{\sigma_{0}^2}| h_{\rm{bu}}+{\bf{h}}^{\rm{H}}_{2u} {\bm{\Theta}_2}{\bf{H}}_{12} {\bm{\Theta}_1} {{\bf{h}}_{b1}}+{\bf{h}}^{\rm{H}}_{1u} {\bm{\Theta}_1} {{\bf{h}}_{b1}}\nonumber\\&+{\bf{h}}^{\rm{H}}_{2u} {\bm{\Theta}_2} {\bf{h}}_{b2} |^2 \nonumber \\ 
		&+\frac{1}{2\rho}\sum\limits_{k = 1}^K\left|u_{k}-{ {{\bf{d}}^{\rm{H}}_{k}}{\bf{A}}_{k}{{\bf{w}}_{k}}}+\rho \lambda_{k,1}\right|^2\nonumber \\ 
		&+\frac{1}{2\rho}\sum\limits_{k = 1}^K\left|v_{k}-{ {{\bf{d}}^{\rm{H}}_{k}}{\bf q}}+\rho \lambda_{k,2}\right|^2 \nonumber\\ 
		&+\frac{1}{2\rho}\left|y_{k,m}-{ {{\bf{d}}^{\rm{H}}_{k}}{\bf{A}}_{m}{{\bf{w}}_{k}}}+\rho \lambda_{k,3}\right|^2,
		\\ 
 \textrm{s.t.}\quad \label{CCCP2}
		&	\eta({\left| {y_{k}} \right|^2}+{P^{\rm{pas}}_{\rm{t}}}{\left| {v_{k}} \right|^2}+\sigma^2{{\left\| { {{\bf{d}}^{\rm{H}}_{k}}} \right\|^2_2}})\nonumber\\&\quad-2 \text{Re}((u^{(i)}_k)^*u_k)+{\left|u^{(i)}_k\right|^2} \le 0, \\ 
		&\sum\limits_{k = 1}^K {\left\| {{{\bf{w}}_{k}}} \right\|^2} \le P_{\rm{r}},\\
		& \left|{\bm{\phi}_1}(n)\right|=1, n = 1, \cdots, N_1, \\ 
		& \left|{\bm{\phi}_2}(n)\right|=1, n = 1, \cdots, N_2,
	\end{align}
\end{subequations}
where ${\bf{\Omega}}_2={\left\{{\bf{w}}_k,{\bf{d}}_k, {\bf{\Theta}}_1,{\bf{\Theta}}_2, u_{k}, v_k,y_{k}\right\}}$, ${P^{{\rm{pas}}}_{{\rm{t}}}}$ denote the transmitted power at BS. Since this problem is similar to Problem (\ref{finalproblem}), we can still use the BCD algorithm to optimize each variable separately.
 The optimal solutions of ${\left\{{\bf{w}}_k,{\bf{d}}_k, u_{k}, v_k,y_{k}\right\}}$ for passive RIS have the same form as in the active RIS scenario. Therefore, in the following part, we just discuss how to design the phase shift variable ${\bf{\Theta}}_1$ and ${\bf{\Theta}}_2$ for passive RISs.
\subsubsection{Optimizing Phase Shift ${\bf{\Theta}}_1$}
The original Problem (\ref{passive_problem}) can be rewritten as
\begin{subequations}\label{sub10n}
	\begin{align}
		\mathop {\max }\limits_{{\bm{\phi}}_1} \quad  \label{sub10n_1}
		&{\bm{\phi}^{\rm{H}}_1}{\bm{\Xi}}_{3}{\bm{\phi}_1}-2{\rm{Re}}\{{\bm{\phi}^{\rm{H}}_1}{\bf{s}}_1\},
		\\ 
		\qquad\ \textrm{s.t.}\qquad 
		&\left|{\bm{\phi}_1}(m)\right|=1, m = 1, \cdots, N_1,
	\end{align}
\end{subequations}
where the parameters are defined as follows
\begin{subequations}
\begin{align}
	{\bm{\Xi}}_{3}&={\bf{a}}{\bf{a}}^{\rm{H}}-\frac{1}{2\rho}\sum\limits_{k = 1}^K{\bf{b}}_k{\bf{b}}_k^{\rm{H}},
{\bf{s}}_1=\frac{1}{2\rho}\sum\limits_{k = 1}^K{\bf{b}}_k{\rm{m}}_k-{{a}}_0{\bf{a}},\\ a_0&={h_{bu}+{\bf{h}}^{\rm{H}}_{2u} {\bm{\Theta}_2} {\bf{h}}_{b2}},\\
	{\bf{a}}^{\rm{H}}&=({\bf{h}}^{\rm{H}}_{2u} {\bm{\Theta}_2}{\bf{H}}_{12}+{\bf{h}}^{\rm{H}}_{1u}){\rm{diag}}({{\bf{h}}_{b1}}). 
\end{align}
\end{subequations}

Since Problem (\ref{sub10n}) has unit-modulus constraints, it is a non-convex problem. We can use the MM algorithm to solve this problem \cite{sun2016majorization}.

For any given solution ${\bm\phi}^{{t}}_1$ at the ${{t}}$-th iteration and
for any feasible ${\bm\phi}_1$, we have
\begin{align}\label{trac51}
	{\bm{\phi}^{\rm{H}}_1}{\bm{\Xi}}_{3}{\bm{\phi}_1} \geq &2{\rm{Re}}\left\{{\bm{\phi}^{\rm{H}}_1}({\bm{\Xi}}_{3}-\lambda_{\rm{min}}{\bf{I}}_{N1}){\bm{\phi}}^{{t}}_1\right\}\\&\nonumber+({\bm{\phi}}^{{t}}_1)^{\rm{H}}(\lambda_{\rm{min}}{\bf{I}}_{N1}-{\bm{\Xi}}_{3}){\bm{\phi}}^{{t}}_1+\lambda_{\rm{min}}{N}_1,
\end{align}
where $\lambda_{\rm{min}}$ is the minimum eigenvalue of ${\bm{\Xi}}_3$.

Thus, Problem (\ref{sub10n}) can be transformed into a more tractable form as follows
\begin{subequations}\label{sub12n}
	\begin{align}
		\mathop {\max }\limits_{{\bm{\phi}}_1} \quad  \label{sub12n_1}
		&2{\rm{Re}}\left\{{\bm{\phi}}^{\rm{H}}_1{\bf{v}}^{{t}}_1\right\},
		\\ 
		\qquad\ \textrm{s.t.}\qquad \label{sub12n_2}
		&\left|{\bm{\phi}_1}(n)\right|=1, n = 1, \cdots, N_1,
	\end{align}
\end{subequations}
where ${\bf{v}}^{{t}}_1=({\bm{\Xi}}_{3}-\lambda_{\rm{min}}{\bf{I}}_{N_1}){\bm{\phi}}^{{t}}_1-{\bf{s}}_1$.
\begin{algorithm}[t]
	\caption{{MM Method to Solve Problem (\ref{sub10n})}}\label{iteda3}
	\begin{algorithmic}[1]
		\STATE Initialize ${{t}}=0$, ${\bm{\phi}^{{(t)}}_1}$;
		\REPEAT
		\STATE Compute the value of  ${\bf{v}}^{{t}}_1$;
		\STATE {Update ${\bm{\phi}^{{(t+1)}}_1}=e^{j\angle({\bf{v}}^{{t}}_1)}$ };
		\UNTIL{the objective function converges}.
	\end{algorithmic}
\end{algorithm}

Then, the optimal solution ${\bm{\phi}^{\rm{opt}}_1}$ can be given by
\begin{eqnarray}\label{opt1}
	{\bm{\phi}^{{(t+1)}}_1}=e^{j\angle({\bf{v}}^{{t}}_1)}.
\end{eqnarray}

The overall algorithm to solve Problem (\ref{sub12n}) is summarized in Algorithm \ref{iteda3}.
\subsubsection{Optimizing Phase Shift ${\bf{\Theta}}_2$}
For this subproblem, we can still optimize the phase shift matrix ${\bf{\Theta}}_2$ by using the same method as optimizing the matrix ${\bf{\Theta}}_1$. Therefore, the original Problem (\ref{passive_problem}) can be transformed into
\begin{subequations}\label{sub11n}
	\begin{align}
		\mathop {\max }\limits_{{\bm{\phi}}_2} \quad  \label{sub11n_1}
		&2{\rm{Re}}\left\{{\bm{\phi}}^{\rm{H}}_2{\bf{v}}^{{t}}_2\right\},
		\\ 
		\qquad\ \textrm{s.t.}\qquad \label{sub11n_2}
		&\left|{\bm{\phi}_2}(n)\right|=1, n = 1, \cdots, N_2,
	\end{align}
\end{subequations}
where the parameter can be defined as follows
\begin{subequations}
	\begin{align}
		{\bf{v}}^{{t}}_2&=({\bm{\Xi}}_{4}-\lambda_{\rm{min}}{\bf{I}}_{N_2}){\bm{\phi}_2^{{t}}}-\sum\limits_{k =1}^K\frac{1}{2\rho}{\bf{r}}_k{\rm{n}}_k+{{b}}_0{\bf{b}}, \\
		{\bm{\Xi}}_{4}&={\bf{b}}{\bf{b}}^{\rm{H}}-\frac{1}{2\rho}\sum\limits_{k = 1}^K{\bf{R}}_k, 	 b_0={h_{bu}+{\bf{h}}^{\rm{H}}_{1u} {\bm{\Theta}_1} {\bf{h}}_{b1}},\\
		{\bf{b}}^{\rm{H}}&={\bf{h}}^{\rm{H}}_{2u} ({\rm{diag}}({\bf{H}}_{12} {\bm{\Theta}_1} {{\bf{h}}_{b1}})+{\rm{diag}}( {{\bf{h}}_{b2}}))
	.
	\end{align}
\end{subequations}

Finally, the optimal solution ${\bm{\phi}^{\rm{opt}}_2}$ can be given by
\begin{eqnarray}\label{opt2}
	{\bm{\phi}^{({{t}}+1)}_2}=e^{j\angle({\bf{v}}^{{t}}_2)}.
\end{eqnarray}
\section{Simulation Results}\label{simu}
In this section, numerical results are provided to demonstrate the advantages of using double active RISs compared with single active RIS and double passive RISs in RCC systems.
\begin{figure}
	\centering
	\includegraphics[width=0.75\linewidth]{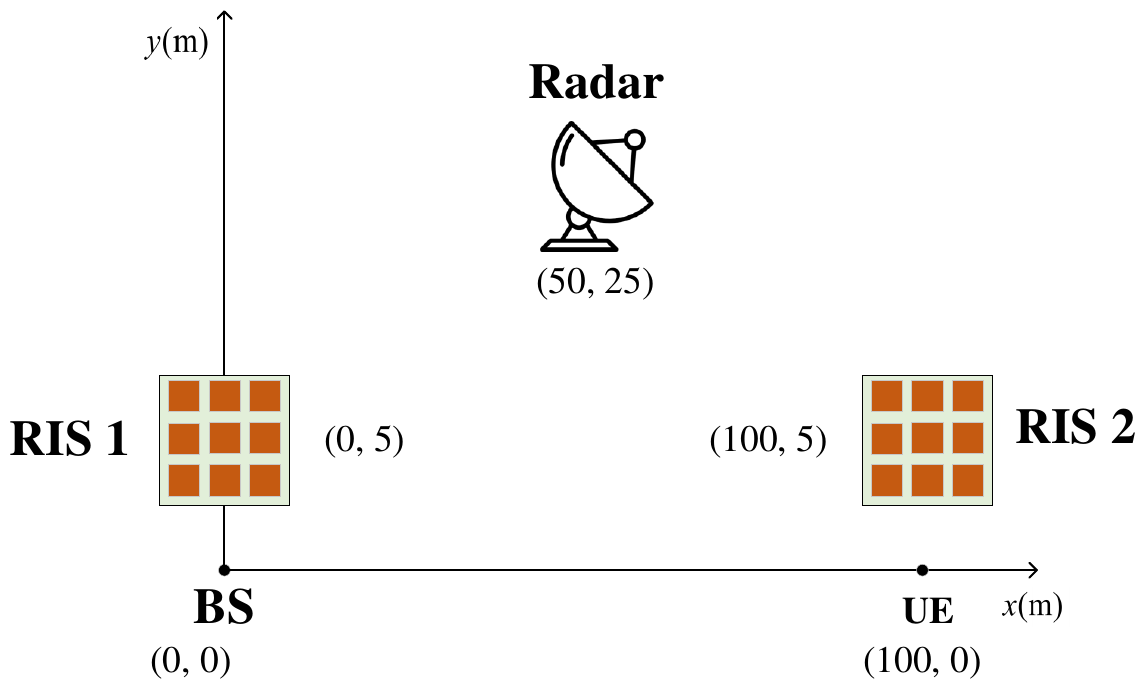}
	\caption{{Simulation proposition.}}\vspace{-0.5cm}
	\label{fig3}
\end{figure}
\subsection{Simulation Parameter Setting}
As shown in Fig. \ref{fig3}, we consider a double RIS-assisted RCC system with two-dimensional coordinates where the location of BS, UE, active RIS 1, active RIS 2, and radar are set to (0 m, 0 m), (100 m, 0 m), (0 m, 5 m), (100 m, 5 m), and (50 m, 25 m), respectively. 
Both BS and UE are equipped with a single antenna. The number of reflecting elements of two RISs are $N_1=N_2=40$. The number of antennas for the radar is $M=12$, and there are $K=7$ directions which is $[-\pi /2,\text{ }-\pi /3,\text{ }-\pi /6,\text{ }0,\text{ }\pi /6,\text{ }\pi /3,\text{ }\pi /2]$. The pathloss factors of target response matrix are ${{\alpha }_{k}}=10^{-1}, \forall k \in \mathcal{K}$. In addition, the radar SINR requirement is set to $\eta =20\text{ dB}$.

The large-scale path loss of the channel in dB is modelled as
 \begin{equation}\label{shqidh}
{\rm{PL = }}{{\rm{P}}{{\rm{L}}_0} - 10\alpha {{\log }_{10}}\left( {\frac{d}{{{d_0}}}} \right)},
\end{equation}
where $\alpha$ is the large-scale path-loss factor, ${\rm{P}}{{\rm{L}}_0}$ is the path-loss value when the reference distance is $d_0$, and $d$ is the distance between the transmitter and the receiver. In our simulations, we assume $d_0=1$ m, and ${\rm{P}}{{\rm{L}}_0}=-30$ dB. 
Due to the obstacles and scatterers, we assume the path loss factors for channel $h_{bu}$, ${\bf{h}}_{b1}$, ${\bf{h}}^{\rm{H}}_{2u}$, ${\bf{h}}_{b2}$, ${\bf{h}}^{\rm{H}}_{1u}$, ${\bf{h}}_{br}$, ${\bf{H}}_{1r}$, ${\bf{H}}_{2r}$, and ${\bf{H}}_{12}$ are ${{\alpha }_{\textrm{BS-UE}}}=3.75$, ${{\alpha }_{\textrm{BS-RIS1}}}={{\alpha }_{\textrm{RIS2-UE}}}=2.5$, ${{\alpha }_{\textrm{BS-RIS2}}}={{\alpha }_{\textrm{RIS1-UE}}}=3$, and ${{\alpha }_{\textrm{BS-Radar}}}={{\alpha }_{\textrm{RIS1-Radar}}}={{\alpha }_{\textrm{RIS2-Radar}}}={{\alpha }_{\textrm{RIS1-RIS2}}}=2.2$, respectively.

The small-scale fading which follows Rician fading can be modeled as
\begin{equation}\label{fvgbhnjik}
	{{\bf{\tilde H}}} = \sqrt {\frac{{{\beta}}}{{{\beta} + 1}}} {\bf{\tilde H}}^{{\rm{LoS}}} + \sqrt {\frac{1}{{{\beta} + 1}}} {\bf{\tilde H}}^{{\rm{NLoS}}},
\end{equation}
where $\beta$ is the factor of Rician fading, which is assumed to $3$. ${\bf{\tilde H}}^{{\rm{LoS}}}$ and ${\bf{\tilde H}}^{{\rm{NLoS}}}$  respectively denote the line of sight (LoS) and the non-LoS (NLoS) channel which follows Rayleigh fading.  ${\bf{\tilde H}}^{{\rm{LoS}}}$ can be modeled as ${\bf{\tilde H}}^{{\rm{LoS}}} = {{\bf{a}}_{{D_r}}}\left( {\vartheta^{AoA}} \right){\bf{a}}_{D_t}^{\rm{H}}\left( {\vartheta^{AoD}} \right)$, where \begin{equation}\label{tgthbty}
\scriptstyle
	{{\bf{a}}_{{D_r}}}\left( {\vartheta^{AoA}} \right) = {\left[ {1,{e^{j\frac{{2\pi d}}{\lambda }\sin \vartheta^{AoA}}}, \cdots ,{e^{j\frac{{2\pi d}}{\lambda }({D_r} - 1)\sin \vartheta^{AoA}}}} \right]^{\rm{T}}},
\end{equation}
and
	\begin{equation}\label{jpjookko}
\scriptstyle		{{\bf{a}}_{D_t}}\left( {\vartheta^{AoD}} \right) = {\left[ {1,{e^{j\frac{{2\pi d}}{\lambda }\sin \vartheta ^{AoD}}}, \cdots ,{e^{j\frac{{2\pi d}}{\lambda }(D_t - 1)\sin \vartheta^{AoD}}}} \right]^{\rm{T}}},
	\end{equation}
where  $\lambda$, $d$, $\vartheta^{AoA}$, and $\vartheta^{AoD}$ represent the wavelength, the separating distance of antennas, the arrival angle, and departure angle, respectively. Both $\vartheta^{AoA}$ and $\vartheta^{AoD}$ follow a random distribution within $[0,2\pi ]$. $D_t$ and $D_r$ denote the number of antennas/reflecting elements of RIS in the transmitter and the receiver, respectively. We assume $d/\lambda=1/2$. The noise power is ${{\sigma }^{2}}={{\sigma }_{0}}^{2}={{\sigma }_{1}}^{2}={{\sigma }_{2}}^{2}=-80\text{ dBm}$, and the tolerance error is $\varepsilon =10^{-3}$. 
\subsection{Benchmark Schemes}
To verify the advantages of deploying double active RISs in the RCC system, we compare the proposed algorithm with the following benchmarks:
\begin{itemize}
	\item  \textbf{Double passive RISs}: Two passive RISs are deployed in the RCC system. The corresponding algorithm in this scheme is based on Section IV.
	\item  \textbf{Single active RIS $i$}: Single active RIS $i$ is deployed in the RCC system. For the case of deploying single active RIS 1 near the BS, we set ${\bf{H}}_{12}={\bf{0}}$, ${\bf{H}}_{2r}={\bf{0}}$, ${\bf{h}}_{b2}={\bf{0}}$, and ${\bf{h}}^{\rm{H}}_{2u}={\bf{0}}$ and skip the step of optimizing ${\bf{\Theta}}_2$ in the proposed algorithm. For the case of deploying single active RIS 2 near the UE, we set ${\bf{H}}_{12}={\bf{0}}$, ${\bf{H}}_{1r}={\bf{0}}$, ${\bf{h}}_{b1}={\bf{0}}$, ${\bf{h}}^{\rm{H}}_{1u}={\bf{0}}$, and skip the step of optimizing ${\bf{\Theta}}_1$ in the proposed algorithm.
	\item \textbf{No RIS}: RIS is not deployed in the RCC system.  Specifically, we set ${\bf{H}}_{1r}={\bf{0}}$, ${\bf{H}}_{2r}={\bf{0}}$, ${\bf{H}}_{12}={\bf{0}}$, ${\bf{h}}^{\rm{H}}_{2u}={\bf{0}}$, and ${\bf{h}}^{\rm{H}}_{1u}={\bf{0}}$. Then, we skip the steps for optimizing ${\bf{\Theta}}_1$ and ${\bf{\Theta}}_2$ in the proposed algorithm.
\end{itemize}

Then, based on the power model in \cite{zhi2022active}, the total power consumption models for the above schemes are described respectively as follows
\begin{subequations}
	\begin{align}
		{Q}^{\rm{DAR}}_{\rm{total}} &= P^{\rm{act}}_{\rm{t}}+P_{\rm{r}}+P_{1}
		+P_{2}	\nonumber\\&\quad+(N_1+N_2)(P_{\rm{SW}}+P_{\rm{DC}}),\\
		{Q}^{\rm{DPR}}_{\rm{total}} &= P^{\rm{pas}}_{\rm{t}}+P_{\rm{r}}+(N_1+N_2)P_{\rm{SW}},\\
		{Q}^{\rm{SAR}}_{\rm{total}} &= P^{\rm{act}}_{\rm{t}}+P_{i}+P_{\rm{r}}\\&\quad+N_i(P_{\rm{SW}}+P_{\rm{DC}}), \forall i\in \left\{1,2\right\},\\
		{Q}^{\rm{NR}}_{\rm{total}} &= P^{\rm{no}}_{\rm{t}}+P_{\rm{r}},\\
		P_{\rm{r}}&=\gamma Q_{\rm{total}},
	\end{align}
\end{subequations}
where $P_{\rm{SW}}$, $P_{\rm{DC}}$, and $\gamma \in [0,1]$ denote the power consumption of the switch and control circuit for each RIS reflecting element, the direct current bias power for each active reflecting element, and the power allocation factor, respectively.
Unless specified otherwise, we set the same total power consumption, ${Q}_{\rm{total}}=11$ ${\textrm{W}}$ for all schemes. Also, we set $P_{\rm{SW}}=-10{\text{ dBm}}$, $P_{\rm{{DC}}}=-5{\text{ dBm}}$, $P^{\rm{act}}_{\rm{t}}=P_{\rm{1}}=P_{\rm{2}}$, and $\gamma =0.9$. Specifically, we compare the performance difference between the above schemes with the same total power budget and the same total number of reflecting elements of RISs for fairness.

\subsection{Convergence Behavior of the Proposed Algorithm}
\begin{figure}[t]
\centering
\includegraphics[width=0.82\linewidth]{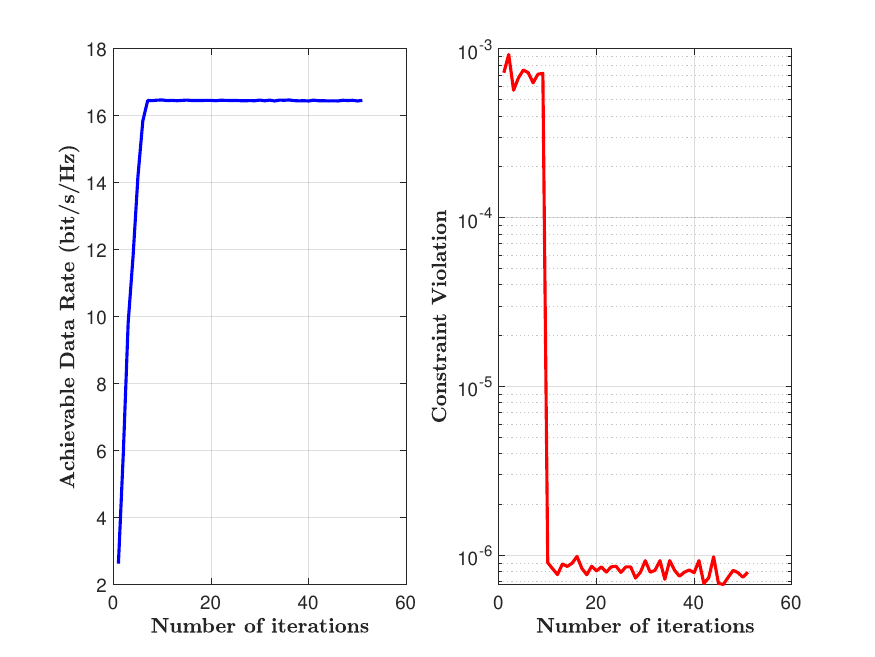}
\caption{{Convergence behavior of PDD algorithm.}}\vspace{-0.5cm}
\label{simu1}
\end{figure}

In this subsection, we set $P_{\rm{r}} = 10$ $\textrm{W}$, $P^{\rm{act}}_{\rm{t}}=P_{\rm{1}}=P_{\rm{2}}= 0.4$ $\textrm{W}$. Fig. \ref{simu1} illustrates the convergence behavior of the proposed algorithm. It is shown that the algorithm quickly converges within about 10 iterations and the value of the constraint violation reaches the level of $10^{-6}$, which allows the equation constraint to hold and ensure the feasibility of the PDD algorithm.

\subsection{The Impact of the Number of Reflecting Elements}
\begin{figure}[t]
	\centering
	\includegraphics[width=0.82\linewidth]{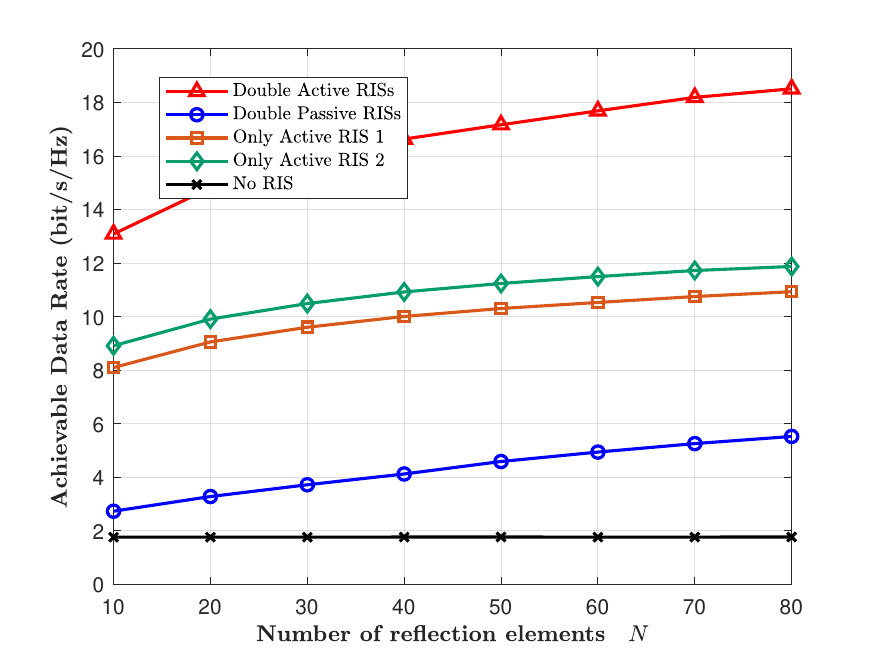}
	\caption{Achievable data rate versus the total number of reflecting elements.}\vspace{-0.4cm}
	\label{simu2}
\end{figure}

Fig. \ref{simu2} shows the achievable data rate versus the number of reflecting elements. Note that the total number of elements $N=N_1+N_2$ is the same for different schemes. As shown in Fig. \ref{simu2}, 
the achievable data rate of the double active RIS-assisted system significantly outperforms the other benchmarks. It illustrates that active RIS can effectively overcome the ``multiplicative fading" effect introduced by passive RIS. Furthermore, the double active RIS-assisted RCC system performs better than the single active RIS-assisted RCC system. This illustrates that deploying double active RISs in RCC system can yield cooperative benefits, thereby enhancing the achievable data rate.
\subsection{The Impact of the Location of RIS}
\begin{figure}[t]
	\centering
	\includegraphics[width=0.82\linewidth]{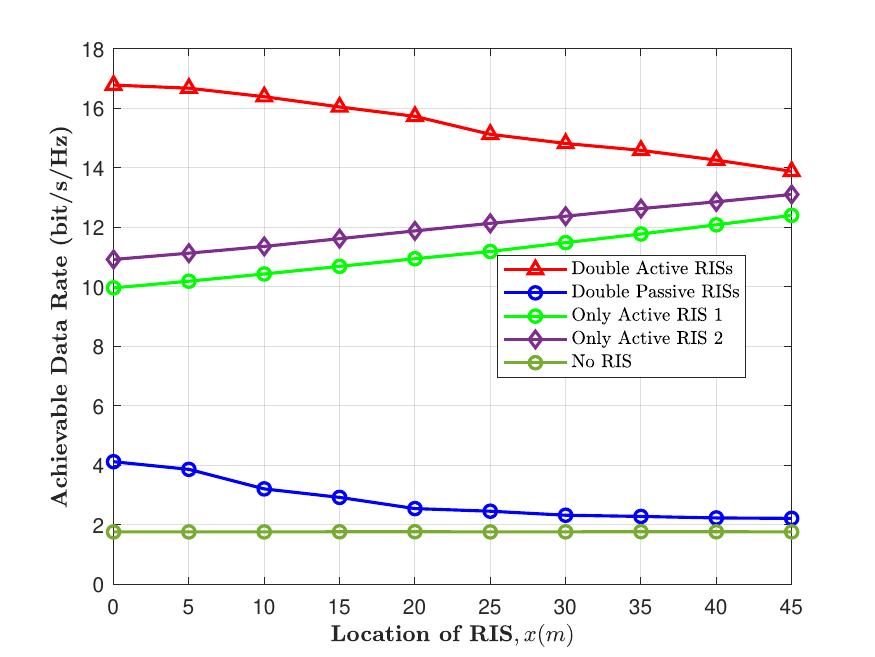}
	\caption{Achievable data rate versus locations of RIS, $x (m)$. }\vspace{-0.4cm}
	\label{simu3}
\end{figure}
In Fig. \ref{simu3}, we investigate the impact of the locations of two RISs on the achievable data rate by moving the positions of two RISs. By denoting the horizontal coordinate of the RIS 1 as $x$, the locations of RIS 1 and RIS 2 are set to ($x$ m, 0 m), ($100-x$ m, 0 m), respectively. It can be seen that the double active RIS-assisted system achieves superior performance over other schemes in all locations. It is interesting to observe that, as $x$ increases, the achievable data rate of the double active RIS-assisted system decreases while data rate of the single active RIS-assisted system increases. This is because as $x$ increase, the power of the signal received by UE decreases due to the increased path loss of the RIS 2-UE link and BS-RIS 1 link.
  Furthermore, the achievable data rate of the two schemes reaches the same level at about $x=45$ m. It shows that the double RISs can be regarded as a single RIS equipped with $N_1+N_2$ reflecting elements as the two RISs get closer to each other.
\subsection{The Impact of the Total System Power}
\begin{figure}[t]
	\centering
	\includegraphics[width=0.83\linewidth]{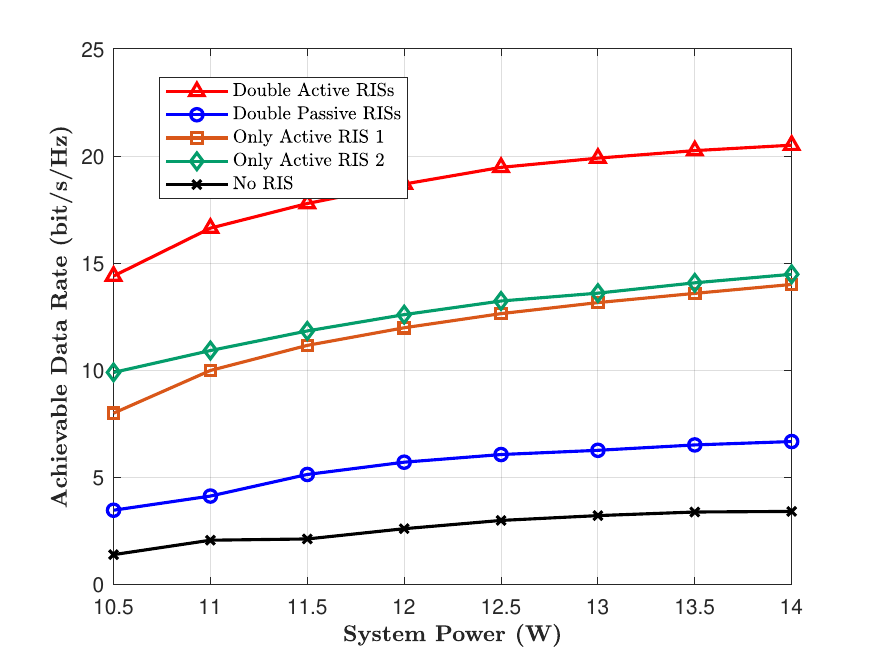}
	\caption{{Achievable data rate versus total system power.}}\vspace{-0.5cm}
	\label{simu4}
\end{figure}
\begin{figure}[t]
	\centering
	\includegraphics[width=0.83\linewidth]{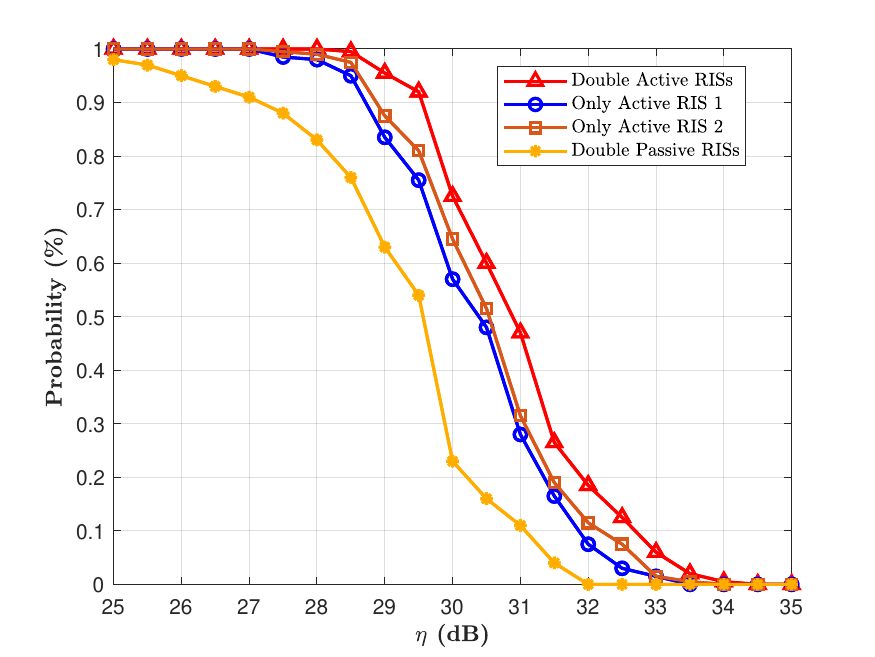}
	\caption{{Successful probability of a feasible solution versus $\eta$.}}\vspace{-0.4cm}
	\label{simu7}
\end{figure}
\begin{figure}[t]
	\centering
	\includegraphics[width=0.83\linewidth]{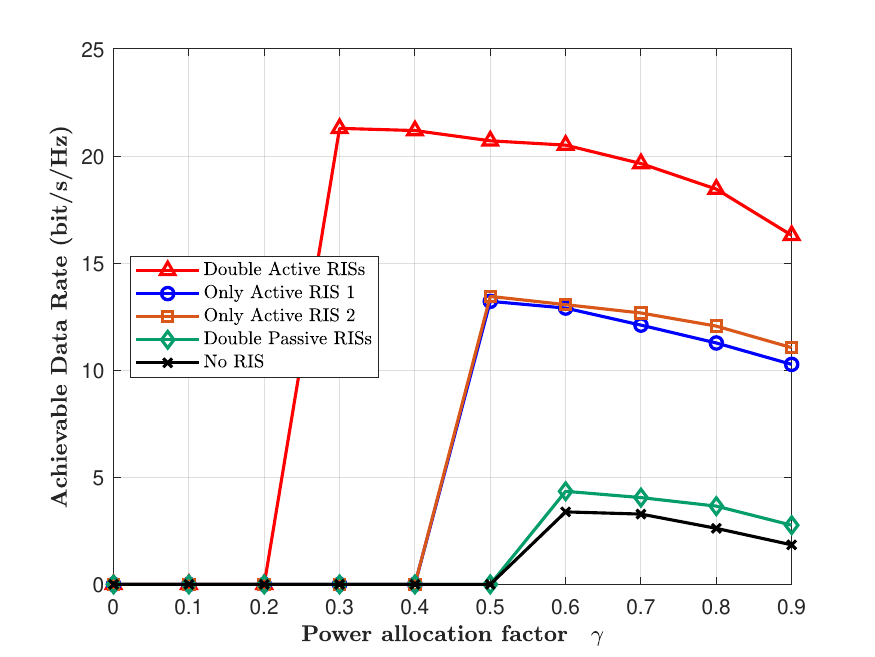}
	\caption{{Achievable data rate versus $\gamma$.}}
	\label{simu5}\vspace{-0.5cm}
\end{figure}
In Fig. \ref{simu4}, we compare the achievable data rate versus the total system power budget. It can be seen that the double active RIS-assisted RCC system exhibits better system performance than the other benchmarks, as the total system power budget increases.  Thus, deploying double active RISs in RCC systems can effectively reduce system power consumption while guaranteeing the required achievable data rate.
\subsection{The Successful Probability of a Feasible Solution versus $\eta$}
In Fig. \ref{simu7}, we study the successful probability of a feasible solution versus radar SINR requirements, i.e., $\eta$. As $\eta$ increases, it becomes more difficult to find a feasible solution that satisfies radar SINR requirements. When $\eta=29$ $ {\rm{dB}}$, the double active RIS-assisted RCC system exhibits a probability of over 95$\%$ to find a feasible solution, whereas the double passive RIS-assisted RCC system has only about a 60$\%$ probability of finding a feasible solution. This clearly shows that interference between the radar and the communication system limits the improvement of radar detection performance i.e., the radar SINR requirement. It demonstrates that double active RISs can more effectively reduce the interference from communication systems to radar compared with other schemes, and thus can improve the detection performance of the radar.
\subsection{The Impact of Power Allocation Factor $\gamma$}
In Fig. \ref{simu5}, we investigate the achievable data rate versus power allocation factor, i.e., $\gamma$. It can be seen that when $\gamma \leq 0.2$, all the schemes fail to find a feasible solution. As $\gamma$ increases, double active RIS-assisted system is the first scheme to find a feasible solution compared with the other schemes,  which means that the double active RISs-assisted RCC system only requires a low radar power budget, i.e., a small value of $\gamma$, to satisfy the radar detection performance i.e., minimum SINR requirement.  
This is because that the double active RIS-assisted RCC system can well suppress the interference from the communication system to the radar.
Furthermore, in the double active RIS-assisted RCC system, radar SINR requirements can be satisfied by assigning less power to the radar under a specified total power budget. Consequently, more power can be allocated to double RISs and BS to enhance the achievable data rate.
\section{Conclusions}\label{conclu}
In this paper, we investigated the jointly beamforming design for a double active RIS-assisted RCC system. We formulated an achievable data rate maximization problem with the guaranteed radar SINR requirement and the power budget limitations of the radar and two active RISs. Because of the nonconvexity of the original problem, we proposed the PDD algorithm based on the FP method and CCP method to convert the original problem to a convex problem. In the inner loop of the PDD framework, we used the BCD method to divide the original problem into several subproblems. Then, we solved the subproblems using the Lagrange dual method. In the outer loop, we updated the dual variables and penalty factor. Simulation results verified the effectiveness of the proposed scheme. We also validated that the proposed algorithm based on the double active RIS-assisted system achieves better system performance than other benchmark schemes under the same power budget and number of reflecting elements. Furthermore,  it is shown that double active RIS-assisted system has superior performance over other schemes in all locations of RISs. 

For simplicity, self-interference between the two active RISs was not considered in this paper, and thus the effect of self-interference on RCC system will be further explored in future research work.

%
%

\





\vspace{-0.5cm}
\bibliographystyle{IEEEtran}
\bibliography{myref}


\end{document}